\documentclass[
twocolumn,superscriptaddress,preprintnumbers,amsmath,amssymb,showpacs,prx,longbibliography]{revtex4-2}
\usepackage{graphicx}
\usepackage{dcolumn}
\usepackage{bm}
\usepackage{amsmath}
\usepackage{amssymb}
\usepackage{color}
\usepackage[dvipsnames]{xcolor}
\usepackage{hyperref}
\hypersetup{
    colorlinks=true,
    linkcolor=,
    filecolor=black,      
    urlcolor=blue,
    citecolor=blue,
}
 \usepackage{natbib}
\usepackage{braket}
\usepackage{soul}
\usepackage{CJK}
\usepackage{relsize}
\usepackage{printlen}
\usepackage{pifont}

\definecolor{testcolor}{rgb}{1    0.5    0}

\newcommand{\yiqiu}{\textcolor{black}}

\begin{document}

\title{Microscopic Reversibility and Emergent Elasticity in Ultrastable Granular Systems}

\author{Yiqiu Zhao}
\email{yiqiuzhao@ust.hk}
\affiliation{Department of Physics, Duke University, Durham, NC, 27708}
\affiliation{Department of Physics, The Hong Kong University of Science and Technology, Hong Kong SAR, China}

\author{Yuchen Zhao}
\affiliation{Department of Physics, Duke University, Durham, NC, 27708}
\affiliation{School of Mechanical and Aerospace Engineering, Nanyang Technological University, 639798, Singapore}

\author{Dong Wang}
\affiliation{Department of Physics, Duke University, Durham, NC, 27708}
\affiliation{Department of Mechanical Engineering \& Materials Science, Yale University, New Haven, CT, 06520}

\author{Hu Zheng}
\affiliation{Department of Physics, Duke University, Durham, NC, 27708}
\affiliation{Department of Geotechnical Engineering, College of Civil Engineering, Tongji University, Shanghai, China}

\author{Bulbul Chakraborty}
\affiliation{Martin Fisher School of Physics, Brandeis University, Waltham, MA, 02454}

\author{Joshua~E.~S.~Socolar}
\email{socolar@duke.edu}
\affiliation{Department of Physics, Duke University, Durham, NC, 27708}

\date{\today}


\begin{abstract}
In a recent  paper~\citep{zhao2022_prx}, 
we reported experimental observations of ``ultrastable'' states in a shear-jammed granular system subjected to small-amplitude cyclic shear.  In such states, all the particle positions and contact forces are reproduced after each shear cycle so that a strobed image of the stresses and particle positions appears static. In the present work, we report further analyses of data from those experiments
\yiqiu{to characterize both global and local responses of ultrastable states within a shear cycle}, not just the strobed dynamics.
\yiqiu{We find that ultrastable states follow a power-law relation between shear modulus and pressure with an exponent $\beta\approx 0.5$, reminiscent of critical scaling laws near jamming.}
We also examine the evolution of contact forces \yiqiu{measured using photoelasticimetry}. We find that there are two types of contacts: non-persistent contacts that reversibly open and close; and persistent contacts that never open \yiqiu{and display no measurable sliding}. We show that the non-persistent contacts make a non-negligible contribution to the emergent shear modulus. We also analyze the  spatial correlations of the stress tensor and compare them to the  predictions of a recent theory of the emergent elasticity of granular solids,
\yiqiu{the Vector Charge
Theory of Granular mechanics and dynamics} (VCTG)~\citep{nampoothiri2020_prl}. 
We show that our experimental results can be fit well by VCTG,  assuming uniaxial symmetry of the contact networks. The fits reveal that the response of the ultrastable states to additional applied stress is substantially more isotropic than that of the original shear-jammed states. Our results provide important insight into the mechanical properties of  frictional granular solids created by shear.

\end{abstract}

\maketitle


\section{Introduction}

Granular materials are athermal collections of  particles that interact with each other only when they form direct, frictional contacts. 
These materials can jam into solid packings that statically resist applied stresses~\cite{Liu1998_nature,Majmudar2007_prl,Hecke2009_jpcm,sarkar2013_prl,luding2016_nature,behringer2018_rpp}. Shear-induced jamming occurs in a variety of disordered, complex systems, including granular \yiqiu{suspensions~\cite{Peters2016_nat,han2019_prl,morris2020_arfm} and dry granular} materials \yiqiu{with~\cite{Bi2011_nat,vinutha2016_np,dong2018_prl,zhao2019_prl,vinutha2019_pre,otsuki2020_pre} or without friction~\cite{Kumar2016_gm,Baity2017_jps,Babu_2021Soft}.}
The stability of shear-jammed states, however, remains
at best partially understood. 
In \yiqiu{a} recent 
\yiqiu{experiment}~\cite{zhao2022_prx}, 
\yiqiu{the stability of shear-jammed states 
\yiqiu{in a frictional granular system}
was systematically examined by monitoring their evolution under small-amplitude, volume-conserving cyclic shear. 
\yiqiu{Many} shear-jammed packings relaxed to a stress-free, diffusive steady state under cyclic strain amplitude as small as 1\%. However, in some cases, the shear-jammed system}
relax\yiqiu{ed} 
into an
\yiqiu{unexpected}
state in which all microscopic degrees of freedom, including particle positions, orientations, and contact forces, remain the same for thousands of shear cycles. 
\yiqiu{These states were termed ``ultrastable'' to distinguish them \yiqiu{from originally formed shear-jammed states that would deform plastically under a single shear cycle} with same strain amplitude. They emerged in athermal, frictional granular packings and are thus qualitatively different from \yiqiu{the states of glasses obtained using vapor deposition that have also been termed ultrastable~\cite{Ediger2017_JCP}.} 
\yiqiu{Nevertheless, the two systems are similar in that} the ultrastable shear-jammed granular packings have smaller pressure and behave more like an elastic ordinary solid than other shear-jammed packings, and ultrastable glasses have lower energy and are more stable against shearing than ordinary glasses~\cite{Ediger2017_JCP,Kim2022_PRL}.}

\yiqiu{In our \yiqiu{earlier work~\cite{zhao2022_prx}, we found that}
a reversibility transition and a jamming/unjamming transition coincide at the phase boundary between the two types of nonequilibrium steady states induced by cyclic shearing: the ultrastable states \yiqiu{that return to the same microscopic configuration after each cycle} and the fluid-like unjammed states in which particles undergo diffusive displacements. 
Without changing the volume fraction, the different types of steady states can be realized by changing either the shear strain $\gamma_{\rm I}$ used to form an original shear-jammed state or the cyclic strain amplitude $\delta\gamma$. A stability diagram is given in Ref.~\cite{zhao2022_prx}.
\yiqiu{Notably, ultrastable states formed by larger $\gamma_{\rm I}$ survive under larger $\delta\gamma$.}} 
\yiqiu{The transition from ultrastable states to unjammed states with increasing $\delta\gamma$ or decreasing $\gamma_{\rm I}$ may be viewed as a yielding transition.} \yiqiu{This transition is similar to the oscillatory yielding of amorphous solids~\cite{regev2013_pre,fiocco2013_pre,nagamanasa2014_pre,keim2014_prl,kawasaki2016_pre}, \yiqiu{which is accompanied by a microscopic reversibility transition that can be classified} as an absorbing-state transition~\cite{nagamanasa2014_pre,ness2020_prl,Reichhardt2022_arxiv}. The ultrastable states reported in Ref.~\cite{zhao2022_prx} are both reversible and mechanically stable and are thus similar to the absorbing states of amorphous solids under cyclic shear with a strain amplitude below the threshold for oscillatory yielding~\cite{nagamanasa2014_pre}, but different from other absorbing states that do not}
\yiqiu{have a mechanically stable structure as in the case of dilute suspensions~\cite{corte2008_natphy}. 
The global stress-strain curves for the ultrastable states reported in Ref.~\cite{zhao2022_prx} appear to be highly elastic, but the internal deformation occurring within individual shear cycles was not examined \yiqiu{for evidence of reversible plastic events~\cite{keim2014_prl} or loops in particle trajectories~\cite{Royer2015_pnas,Otsuki2021_EPJE,nagasawa2019_soft}}.}

\yiqiu{In addition to characterizing the grain scale deformations occurring in the ultrastable state, we study the relation between the global elasticity features and the local stresses at the grain scale. A recent theory, termed the Vector Charge Theory of Granular mechanics and dynamics (VCTG)~\cite{nampoothiri2020_prl,nampoothiri2022_arxiv}, suggests a promising approach for relating the global elastic behaviour to features of the contact forces between individual particles.} 
\yiqiu{VCTG is a stress-only framework 
for amorphous solids; it does not rely on a unique reference structure to define strain.}
\yiqiu{This theory} maps the
mechanical response of granular solids to the static, dielectric response of a tensorial electromagnetism with the electric polarizability of the medium mapping to emergent elastic moduli.
VCTG relates the spatial correlations of the stress-tensor to these elastic moduli, which emerge from the underlying contact and force network. 
While previous experiments confirm\yiqiu{ed} some features of the stress correlation functions predicted by the theory~\cite{nampoothiri2020_prl}, there has not been a direct comparison of the elastic constants obtained from fitting the stress correlations to the elastic constants measured from stress-strain curves in experiments.
It is thus of interest to examine our data in the framework of such a theory.

In the present work, we report a detailed analysis of the elastic properties of the previously reported ultrastable shear-jammed states~\cite{zhao2022_prx}. We find that the emergent shear modulus follows a power-law relation with pressure with an exponent consistent with some numerical models. However, the shear response contains a special non-linear feature: there are many contacts that reversibly open and close under low amplitude cyclic shear. These non-persistent contacts contribute a non-negligible portion to the global effective shear modulus. We also examine the relation between the global elastic constants and internal stress correlations predicted by VCTG. The analysis leads to intriguing scalings of the emergent elastic properties of the system and uncovers a feature that reflects how cyclic shear modifies the elastic properties of a jammed packing. Our results bring new insights to the elasticity of frictional granular materials near jamming.

\begin{figure*}[!t]
    \centering
    \includegraphics[width =\textwidth]{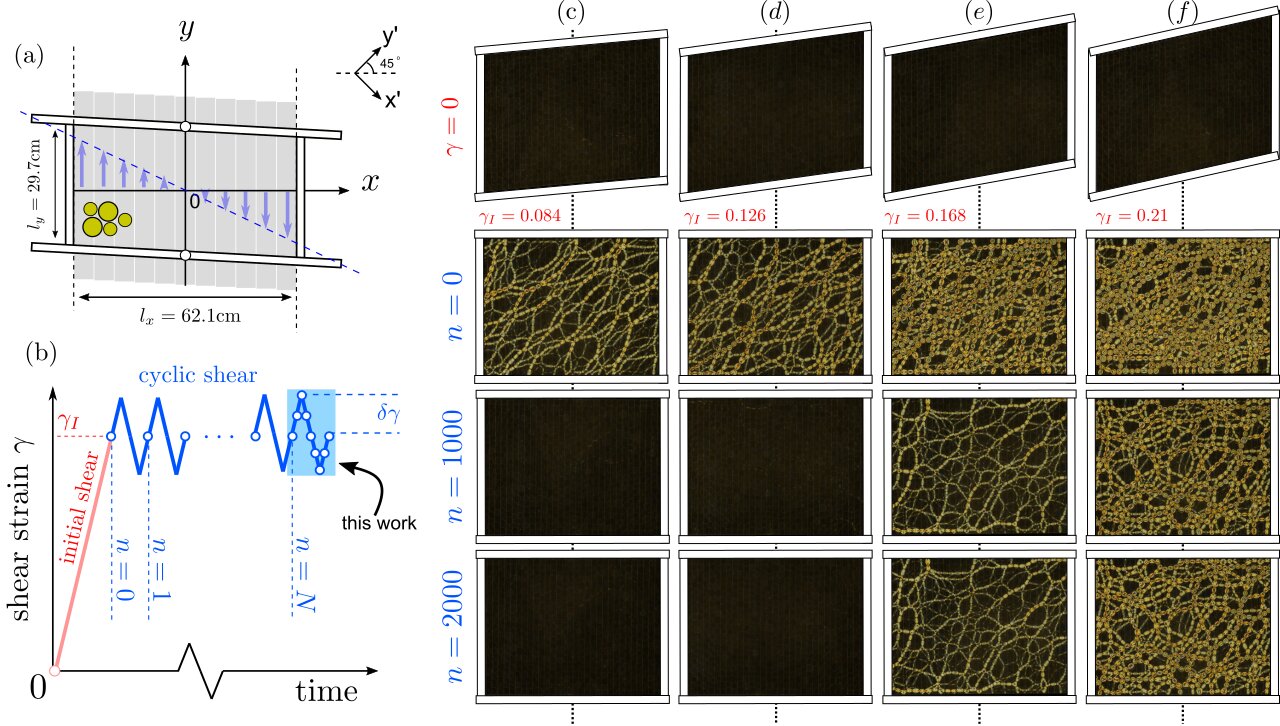}
    \caption{{\bf Experimental protocol and the ultrastable states.} (a) A schematic top view of the multi-slot shear cell.
    \yiqiu{Bottom slots move together with boundary walls to impose a uniform simple shear profile.}
    (b) The applied strain as a function of time. An initial large forward shear is followed by multiple periods of small-amplitude cyclic shear. The shear rate is always in the quasistatic regime. 
    \yiqiu{The light blue rectangle indicates the period of interest for the present work. Open circles schematically indicate the times of data snapshots. Both (a) and (b) are adapted from Ref.~\cite{zhao2022_prx}.
    }
    (c-f) Snapshots of the force chain network for four independent runs with different $\gamma_{\rm I}$. Each column shows images in the original configuration, the shear-jammed configuration following the initial forward shear, and the configurations reached after 1000 and 2000 shear cycles. As indicated by the shapes of the original configurations, the initial forward shear used to reach the rectangular configuration increases from left to right. All systems in (c-f) have same packing fraction $\phi=0.816$, and the amplitudes of the cyclic shear are the same $\delta\gamma = 0.95\%$. The images are taken through a polariscope and thus only particles that bear finite stress are visible. 
    \yiqiu{Two ultrastable states are formed in (e) and (f) as the system locks in jammed states that do not change over at least a thousand shear cycles.}
    } 
    \label{fig:protocol}
\end{figure*}

\section{Methods}

The analyses in the present paper are performed on the same set of experiments reported in Ref.~\cite{zhao2022_prx}. 
\yiqiu{In this work, we focus on the evolution of the system within several shear cycles after an ultrastable state is formed while Ref.~\cite{zhao2022_prx} focused on the strobed states.}
The materials and experimental protocols are briefly summarized here. More details can be found in Ref.~\cite{zhao2022_prx}. 

Our model granular system consists of a bidisperse layer of photoelastic disc with same height, 6.8~mm, but different diameters: $d_{\rm b} = 15.9$ mm $d_{\rm s} = 12.7$ mm. The static friction coefficient between the particles is $\yiqiu{\mu_{\rm s}} = 0.87\pm 0.03$.
\yiqiu{Under static diametric loading,
the normal contact force law is roughly Hertzian. For a
small disc squeezed between to rigid surfaces, we measure
\begin{equation}\label{eq:force_law}
 f_{\rm n} \approx \frac{\epsilon_{\rm s}}{r_{\rm s}}(\delta/d_{\rm s})^{3/2}   
\end{equation}
where $f_{\rm n}$ is the normal contact force, $\delta/d_{\rm s}$ is the diametric strain, $r_{\rm s}$ is the radius of the small disc, and $\epsilon_{\rm s}$ = 2.73 N$\cdot$m. (This expression slightly overestimates the weak forces. Details on contact force law calibration are given in Appendix~A of Ref.~\cite{zhao2022_prx}.)}
The discs are placed in a simple shear box with a parallelogram boundary and a multi-slat base that promotes homogeneous shear when the angle between the boundaries changes. A schematic of the shear apparatus is shown in Fig.~\ref{fig:protocol}(a), and more details can be found in Refs.~\cite{Ren2013_prl,ren2013_thesis}.  The number of particles is fixed at 1040 for all experiments. The area of the shear box is also kept constant throughout. The packing fraction, defined as the total area of particles divided by the area of the shear box, is $\phi=0.816$ for all experiments, which is below the frictionless isotropic jamming point $\phi_{\rm J}\approx 0.835$ estimated using the same apparatus~\cite{Ren2013_prl}.

At the beginning of each experiment, the particles are randomly placed in a zero-stress, unjammed configuration.  When the boundary walls impose a volume-conserving simple shear deformation, the parallel bottom slats move accordingly to impose a uniform internal shear strain field. Static friction causes the particles to move with the slats in an unjammed configuration. Such a substrate-assisted shear protocol avoids boundary-induced density heterogeneity and leads to homogeneous shear-jammed states~\cite{ren2013_thesis}.  The frictional forces between particles and the slats are, however, much smaller in magnitude than the mean contact force in the jammed states.

Starting from the unjammed initial state, each shear experiment consists of two stages: ($i$) an initial shear that forms a shear-jammed state; and ($ii$) a number of consecutive shear cycles that cause the shear-jammed state to transform. The two stages are sketched in Fig.~\ref{fig:protocol}(b). In the initial shear stage, we apply a shear strain $\gamma_{\rm I}$ that transforms an initial chosen parallelogram into a rectangle. Note that the jammed states formed by different $\gamma_{\rm I}$ have different features of the contact and force networks~\cite{Bi2011_nat,sarkar2016_pre,dong2018_prl}. In the cyclic shear stage, we apply a series of $N$ small-amplitude shear cycles with strain amplitude $\delta\gamma \ll \gamma_{\rm I}$. The value of $N$ is at least 1500, and the largest one we used is 4800. In this stage, we only monitor states 
\yiqiu{before and} 
after each complete shear cycle, 
\yiqiu{as sketched by the small circles in Fig.~\ref{fig:protocol}(b)}.
In the last two or three cycles, 
\yiqiu{where an ultrastable state is formed,} we also record images of the system within the shear cycles. 
\yiqiu{The present work focuses on these cycles, as highlighted by the light blue region in Fig.~\ref{fig:protocol}(b).}

At all stages, the imposed shear can be considered quasistatic; the system reaches mechanical equilibrium much faster than the overall shearing rate~\cite{zhao2022_prx}. A high-resolution camera is used to take images of the system, and physical quantities of interest are measured using image processing techniques.
For each state of interest, we measure all the vector contact forces between individual particles using a nonlinear inverse fitting algorithm~\cite{majmudar2005_nature}, details of which can be found in Appendix~C of Ref.~\cite{zhao2022_prx}. The stress tensor is calculated from the measured contact forces as in Refs.~\cite{christoffersen1981_jam,radjai1998_prl,Bi2011_nat} 
\begin{equation}\label{eq:stress_tensor}
    \hat{\sigma} = \frac{1}{S}
    \sum_{i\neq j}^{N_{\rm p}}
    \mathbf{r}_{ij}\otimes \mathbf{f}_{ij},
\end{equation}
where $\mathbf{f}_{ij}$ is the vector force applied to partilce $i$ particle by the particle $j$, $\mathbf{r}_{ij}$ is the 
displacement of the contact point of particles $i$ and $j$ from the center of particle $i$, and the summation runs over particle indices $i$ and $j$
from 1 to the number of particles 
$N_{\rm p}$. We exclude the particles that are in direct contact with the boundary walls, and $S$ in Eq.~\ref{eq:stress_tensor} is the total Voronoi area for the particles that do not belong to the boundary layer.

Ultrastable states are formed when $\gamma_{\rm I}$ is large and $\delta\gamma$ is small.
For smaller $\gamma_{\rm I}$ or larger $\delta\gamma$, the system relaxes to an unjammed, fluid-like state. Fig.~\ref{fig:protocol} shows snapshots illustrating the different behaviors. Figure~\ref{fig:protocol}(c-f) shows example images obtained from typical runs with different $\gamma_{\rm I}$. All of the images are taken through a polariscope so that only the discs supporting finite stress are visible. The second row (labeled $n=0$) shows the stress state after the initial shear $\gamma_{\rm I}$, and the third and fourth rows show the states after 1000 and 2000 shear cycles with $\delta\gamma = 0.95\%$. The nearly blank images in columns (c) and (d) indicate that the system has relaxed to a steady state with nearly zero pressure. The close similarity between images in the third and fourth rows of columns (e) and (f) indicate that ultrastable states are reached within 1000 cycles. Our focus in this paper is on the elastic properties of these ultrastable states.

\section{Results}\label{sec_results}

\subsection{Emergent shear modulus}

\begin{figure*}[!t]
    \centering
    \includegraphics[width =\textwidth]{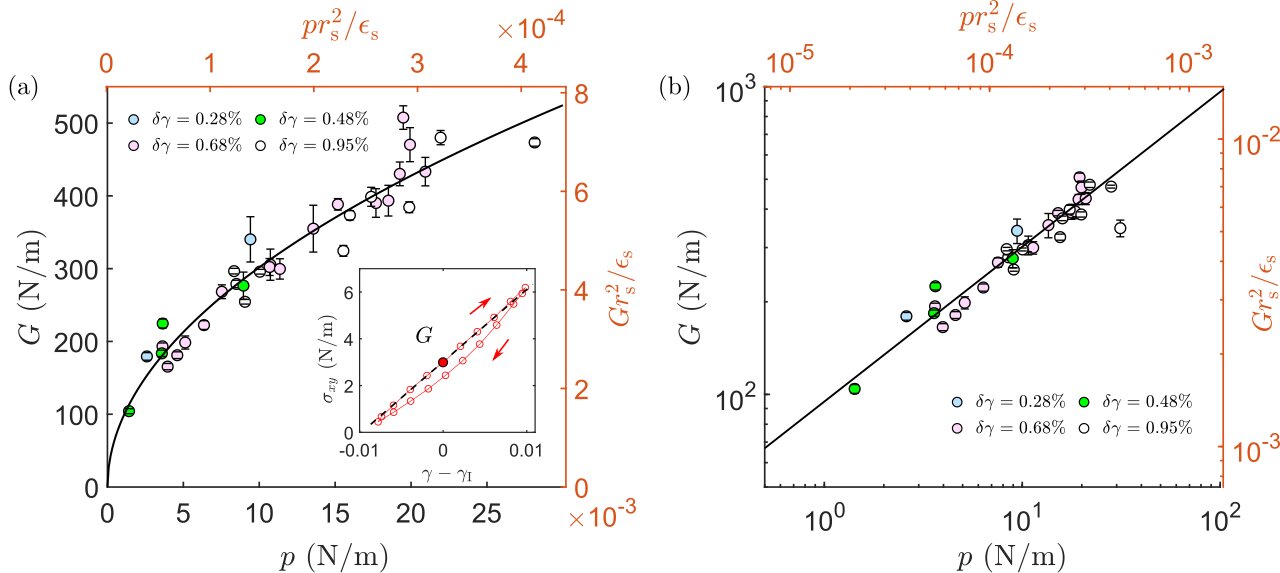}
    \caption{{\bf The emergent shear modulus of ultrastable states formed by different $\gamma_{\rm I}$ and $\delta\gamma$ as a function of pressure.} 
    \yiqiu{(a)} Insert: an example evolution of shear stress $\sigma_{xy}$ versus strain under cyclic shear for an ultrastable state. The shear modulus $G$ is defined as the slope of the forward branch of the curve. Main panel: Measurements of the shear modulus $G$  for the ultrastable states created by cyclic shear with different strain amplitude $\delta\gamma$. Each data point corresponds to an independent packing. 
    The brown axes show the corresponding dimensionless values for pressure and shear modulus
    \yiqiu{where $r_{\rm s}$ and $\epsilon_{\rm s}$ are from Eq.~\ref{eq:force_law}}.  The black curve shows a power-law fit of the form $G\propto p^{\beta}$ with $\beta = 0.5$.
    \yiqiu{(b) Log-log plot of the data as in the main panel of (a).}} 
    \label{fig:stable_modulus}
\end{figure*}

We first examine the global shear modulus $G$ for the ultrastable states under cyclic shear. The insert panel of Fig.~\ref{fig:stable_modulus}(a) plots an example $\sigma_{xy}$ evolution in a shear cycle. The filled red circle is the ultrastable state being considered. \yiqiu{The behavior appears similar to a viscoelastic material in a highly elastic regime. The finite area enclosed by the curve suggests that there is measurable energy dissipation inside the system, although this hysteresis in the stress-strain curve is much smaller than that of shear-jammed states formed by initial shear alone~\cite{zhao2022_prx}. The microscopic mechanisms responsible for this small dissipation could be the sliding of particles over the base and the confining walls or the sliding at inter-particle contacts. We have examined the distribution of the tangential to normal contact force ratio and find that most force-carrying contacts are far from the Coulomb threshold for sliding. However, we cannot exclude the existence of reversible sliding at weak contacts. In addition, the viscoelasticity of the polyurethane photoelastic discs leads to small but measurable hysteresis in the force-displacement curve for a single particle under cyclic diametric loading, as shown in the Appendix A in Ref.~\cite{zhao2022_prx}. This material effect may also contribute to the global hysteresis in the stress-strain curve.}

\yiqiu{From numerical simulations, it is known that both the elastic constants and the stresses of jammed granular materials follow scaling laws in the vicinity of the jamming point~\cite{OHern2003_pre,somfai2007_pre,Goodrich2016_pnas}. 
While the exponents associated with stresses and contact numbers have been examined in experiments~\cite{Majmudar2007_prl}, the scaling of elastic moduli remains largely unexplored, especially for frictional systems. Previous experiments measured the scaling indirectly through acoustic propagation~\cite{makse2004_pre}. Our experimental system 
allows us to study the scaling of the shear modulus of the ultrastable states.} 

We define the shear modulus $G$ as the slope of the curve in the vicinity of the ultrastable state. In practice, we fit a straight line to the rising branch of the curve and obtain the fitted slope. \yiqiu{The jamming point has zero pressure. Thus, the pressure $p$ serves as the measure of distance to the jamming point. While the excess contact number may be a more fundamental quantity, pressure is measured with higher accuracy than contact number in photoelastic experiments. Thus, in this paper, we focus on the relation between $G$ and $p$. We also note that this relation is of great engineering interest~\cite{makse2004_pre}, as in real applications it is easier to control the pressure in a packing than the average contact number.} Figure~\ref{fig:stable_modulus}(a) shows the measured values of $G$ plotted as a function of pressure for various small values of $\delta\gamma$. Without any rescaling, the data falls on a single curve\yiqiu{, suggesting that all the ultrastable states are governed by a universal scaling relation despite the rather special protocol used to generate them}. \yiqiu{The independence of $\delta\gamma$ also suggests that the ultrastable states are below the onset of softening~\cite{Ishima2020_PRE}.}
The solid curve shows a power-law fit of the form
 \begin{equation}\label{eq:G_p_scaling}
 G= G_0 p^{\beta}\,,
 \end{equation} 
 with the fit parameters $G_0=(95\pm 15)\,$N/m and $\beta=0.50\pm 0.06$. 
 \yiqiu{A log-log plot of same data is shown in Fig.~\ref{fig:stable_modulus}(b).}
 Interestingly, the shear modulus measured using sound propagation for isotropic systems exhibits similar dependence on pressure~\cite{makse2004_pre}. Numerical simulations with frictionless spherical particles near jamming interacting through linear spring force laws also show $\beta=0.5$ for both isotropic~\cite{OHern2003_pre,Goodrich2016_pnas,vanderwerf2020_prl} and 
 \yiqiu{shear-jammed} 
 systems~\cite{Baity2017_jps}. 
 For frictionless Hertzian contact models simulations give $\beta=2/3$ over the range of dimensionless pressures studied in our system~\cite{OHern2003_pre,wang2021_pre}. Simulations of 2D packings of frictional spheres with Hertz-Mindlin forces and with friction coefficients similar to ours show $\beta$ between 1/2 and 2/3~\cite{somfai2007_pre}. 
 \yiqiu{As mentioned above, the contact force law between our discs is roughly Hertzian, and our measured $\beta$ appears similar to those obtained in numerical simulations. 
 In addition, the range of dimensionless pressures in our experiments falls in the range studied in simulations, suggesting that our system is indeed close enough to the jamming point to be in the scaling regime.
 We note that a recent simulation of 2D frictional particles under oscillatory shear  with linear-dashpot contact model also found $\beta= 1/2$ for the small-strain plateau shear modulus~\cite{Ishima2020_PRE}. Another recent simulation shows that $\beta$ can be different for deformable particles whose shape is controlled by surface tension rather than internal bulk stresses~\cite{treado2020_arxiv}.}

\subsection{Persistent and non-persistent contacts}

The internal deformation of the system  exhibits non-trivial features. Particle displacements are non-affine, and many force-bearing contacts are activated and deactivated reversibly during a shear cycle. These contacts contribute to the emergent elastic moduli through a nonlinear process that can not be predicted by analyzing a single contact network.

We here report experimental characterizations of the two types of contacts that contribute to the emergent elastic modulus of the packing in a shear cycle. The non-persistent contacts are those that break reversibly during a shear cycle, while the persistent contacts never break once the ultrastable state is reached.

\subsubsection{Non-persistent contacts}

\begin{figure*}[!t]
    \centering
    \includegraphics[width =\textwidth]{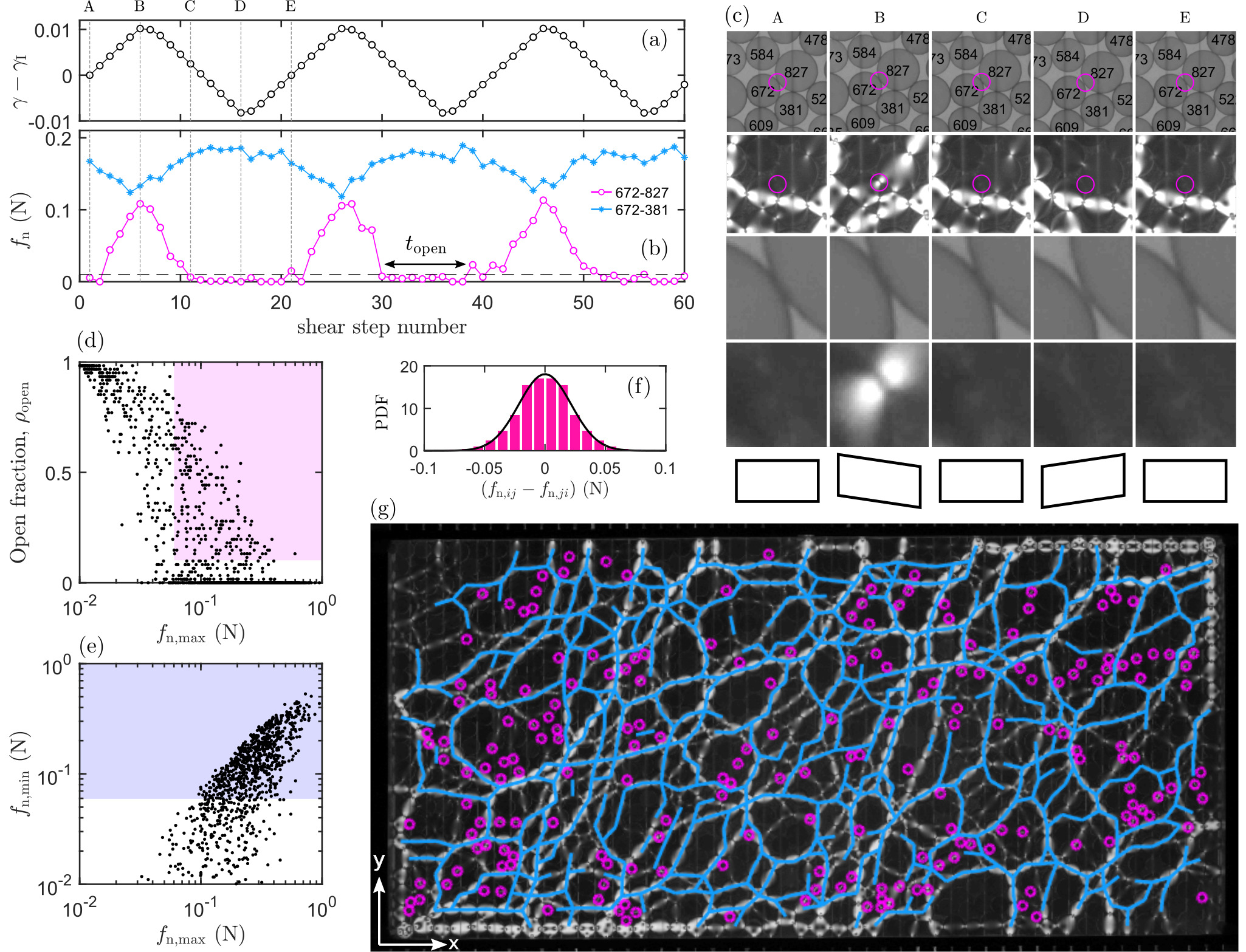}
    \caption{{\bf Non-persistent and persistent contacts.} All data in this figure are from a single example ultrastable state formed by $\gamma_{\rm I} = 14.7\%$ and $\delta\gamma = 0.95\%$. (a) The boundary shear strain as function of shear steps during \yiqiu{three} shear \yiqiu{cycles with same strain amplitude after the ultrastable state is formed (i.e., the strobed states remained unchanged for at least a thousand of cycles before the three cycles shown here)}.
    \yiqiu{Each shear cycle contains 20 shear steps.} (b) The normal force magnitude, $f_{\rm n}$\yiqiu{,} on the contact between particles 672 and 827 (purple circles) and on the contact between particles 672 and 381 (blue stars). The \yiqiu{horizontal}
    dashed line marks a threshold value 0.01 N.
    \yiqiu{The number of shear steps that a contact remains open is noted as $t_{\rm open}$.}
     (\yiqiu{c}) Snapshots of regions from five different states taken from the cycle shown in (a). Each column shows, from top to bottom, a small region of the packing imaged without the polarizer, the same region imaged through the polarizer, a region around a single contact (particles 672 and 827) imaged without the polarizer, and the same region imaged with the polarizer. Note that in column D, contact 672-827 is clearly opened (see the visible gap between particles in the third row), while in B it bears a force $\sim 0.1$ N. The shape at the bottom of each column shows the boundary configuration of the shear cell. See 
    \href{https://drive.google.com/file/d/1xBV0dKmBwXwKUr4TcnNRAzQSTD073u4p/view?usp=sharing}{supplementary video 1} for a video of this process. 
    (\yiqiu{d}) Scatter plot of $\rho_{\rm open}$\yiqiu{, the ratio between $t_{\rm open}$ and the total number of shear steps per cycle,} versus $f_{\rm n.max}$\yiqiu{, the largest $f_{\rm n}$ over a shear cycle,}
    for all contacts. The contacts in the purple region are selected as the non-persistent contacts and are indicated by purple circles in (g). 
    (\yiqiu{e}) Scatter plot of $f_{\rm n,min}$\yiqiu{, the smallest $f_{\rm n}$ over a shear cycle,} versus  $f_{\rm n,max}$ for all contacts.
    The contacts in the blue region     are selected as the persistent contacts and are indicated by blue line segments in (g).
    (\yiqiu{f}) The probability density function for the discrepancy between the measured action and reaction normal forces. The black curve shows a Guassian fit with a width near 0.03 N.
    (g) Image showing the classified contacts. Purple circles mark non-persistent contacts and blue line segments indicate persistent contacts. \yiqiu{Stressed particles that have unclassified contacts are \yiqiu{unmarked but} visible in underlying polarized image \yiqiu{taken at state A as shown in (a)}.} 
    } 
    \label{fig:non_persistent_contact}
\end{figure*}

We first demonstrate the existence of non-persistent contacts in an example ultrastable state formed by initial shear $\gamma_{\rm I} = 14.7\%$ and cyclic shear amplitude $\delta\gamma = 0.95\%$. After the system has settled in the ultrastable state for thousands of cycles, we examine the response of the system during the next three shear cycles. Figure~\ref{fig:non_persistent_contact}(a) plots the global shear strain for these three shear cycles. 

We show that the contact between particles 672 and 827 shown in Fig.~\ref{fig:non_persistent_contact}(c) is a non-persistent contact. The magnitude of the normal component of the contact force, $f_{\rm n}$, on this contact is plotted in Fig.~\ref{fig:non_persistent_contact}(b). It reversibly drops to zero.
To further show that the contact actually opens, we show snapshots of the system in five typical states A to E in Fig.~\ref{fig:non_persistent_contact}(c). In state D, as shown by the snapshot in the third column of Fig.~\ref{fig:non_persistent_contact}(c), the contact is clearly opened. In state B, the clearly visible photoelastic fringes confirm that the contact is carrying finite forces. Thus, this contact opens and closes reversibly in a shear cycle, and is called a non-persistent contact in this paper. See \href{https://drive.google.com/file/d/1xBV0dKmBwXwKUr4TcnNRAzQSTD073u4p/view?usp=sharing}{supplementary video 1} for a video of this process. \yiqiu{Note that the evolution of $f_{\rm n}$ is consistent with the global shear. The branch vector pointing to contact 672-827 from the center of particle 672 is roughly parallel to the $y'$ direction (see Fig.~\ref{fig:protocol}(a)), along which the system is compressed from state A to state B and is stretched from B to D. Accordingly, $f_{\rm n}$ on contact 672-827 grows from A to B and drops from B to D, during which it opens.}

To quantitatively classify the contacts and characterize the behavior of the non-persistent contacts, we consider two characteristic quantities: (1) the fraction of time that the contact is open, $\rho_{\rm open}$, defined as the number of steps in one complete shear cycle for which $f_{\rm n}<0.01$ N divided by the total number of steps in the cycle; and (2) the maximum value of $f_{\rm n}$ for this contact over the whole shear cycle, $f_{\rm n,max}$. Figure~\ref{fig:non_persistent_contact}(d) shows a scatter plot of $f_{\rm n,max}$ and $\rho_{\rm open}$ for all contacts measured over three consecutive shear cycles.

By definition, a non-persistent contact should have a non-zero $\rho_{\rm open}$ and a non-zero $f_{\rm n,max}$. For present purposes, we detect the non-persistent contacts with $\rho_{\rm open}>0.1$ and $f_{\rm n,max}>0.06$ N (the points in the purple region in Fig.~\ref{fig:non_persistent_contact}(d). We intentionally choose a large value for $\rho_{\rm open}$ to be sure the the contact actually opens and a large value for $f_{\rm n,max}$ to ensure that the contact actually closes. We emphasize that the goal here is to demonstrate the existence of non-persistent contacts and their relevance in contributing to the elastic responses of the packing. This conservative classification method ensures that a positive result is meaningful. 

The threshold for $f_{\rm n,max}$ is chosen based on the uncertainty of our force solving algorithm. Figure~\ref{fig:non_persistent_contact}(f) plots the probability density function of the difference between action and reaction contact forces determined by our fitting algorithm for all contacts detected in 61 jammed states over the three shear cycles. The width of this distribution is an estimation of the uncertainty of our force measurements because Newton's third law ensures that these differences must actually be zero. A Gaussian fit gives a width around $0.03$ N. Thus,a contact is convincingly closed at least once in a shear cycle if $f_{\rm n,max}>0.06$ N.

The threshold for $\rho_{\rm open}$ is chosen based on our sampling frequency. For a shear cycle with strain amplitude $\delta\gamma=0.95\%$, there are 20 quasi-static data collection steps per cycle, as shown in Fig.~\ref{fig:non_persistent_contact}(a). Thus, the resolution of $\rho_{\rm open}$ is 1/20. Therefore, we expect that a threshold value of $1/10$ probes contacts that actually opens during a shear cycle. In Fig.~\ref{fig:non_persistent_contact}(d) there appear to be some data points with $\rho_{\rm open}$ between 0 and 1/20 because they are averaged values over three shear cycles.

The detected non-persistent contacts for the example ultrastable state are plotted on top of the photoelastic fringes in Fig.~\ref{fig:non_persistent_contact}(g). These contacts are scattered in space and do not form a percolating network. As a measure of the prevalence of non-persistent contacts, we calculate $f_{\rm npc}$, the number of non-persistent contacts divided by the total number of contacts that were ever closed during a shear cycle. Figure~\ref{fig:non_persistent_contact_statistics}(a) plots $f_{\rm npc}$ as a function of the initial shear strain $\gamma_{\rm I}$ \yiqiu{for ultrastable states formed using same $\delta\gamma=0.95\%$}. \yiqiu{We find that} $f_{\rm npc}$ is larger for smaller $\gamma_{\rm I}$ and can be as large as about 10\%.

\yiqiu{Notably, reversible plastic events observed in frictionless systems (e.g. in two-dimensional foams \cite{Lundberg2008_pre}) also involve reversibly activated inter-particle contacts. A distinction in our frictional system is that there is no obvious T1 event as observed in Ref.~\cite{Lundberg2008_pre}. For example, the reversible activation of contact 672-827 in Fig.~\ref{fig:non_persistent_contact}(c) is not accompanied by a neighbor switching event for the four particle 672, 827, 381 and 584. In other words, there is no obvious local plastic event triggered by an opening of a non-persistent contact. Thus, the particles with non-persistent contacts are not equivalent to bucklers in isostatic frictionless packings~\cite{Charbonneau2015_prl}, where breaking a contact will immediately induce the formation of a new contact.}

\subsubsection{Persistent contacts}

For an ultrastable state, most of the contacts are persistent; they remain closed throughout the shear cycle. 
One example is the contact  between particles 672 and 381, shown in Fig.~\ref{fig:non_persistent_contact}(c). It always bears a finite normal force component $f_{\rm n}$ under cyclic shear, as shown in Fig.~\ref{fig:non_persistent_contact}(b). In practice, we classify a contact as persistent if the minimum normal force during a whole cycle, $f_{\rm n,min}$, is greater than 0.06 N. 
The threshold is chosen as twice our force measurement uncertainty to make sure that $f_{\rm n,min}$ is convincingly larger than 0. Figure~\ref{fig:non_persistent_contact}(e) plots all the contacts for the example ultrastable state in the $f_{\rm n,min}$ and $f_{\rm n,max}$ plane; all the contacts in the light blue region are classified as persistent contacts.

Unlike the scattered non-persistent contacts, the persistent contacts form a percolating network.
Figure~\ref{fig:non_persistent_contact}(g) shows the network formed by the persistent contacts (light blue lines).
Notably, the persistent contact network formed shown in Fig.~\ref{fig:non_persistent_contact_statistics}(g) features large holes reminiscent of the sponge-like structures revealed by rigidity analysis~\cite{liu2021_prl}. 

\begin{figure*}[!t]
    \centering
    \includegraphics[width =\textwidth]{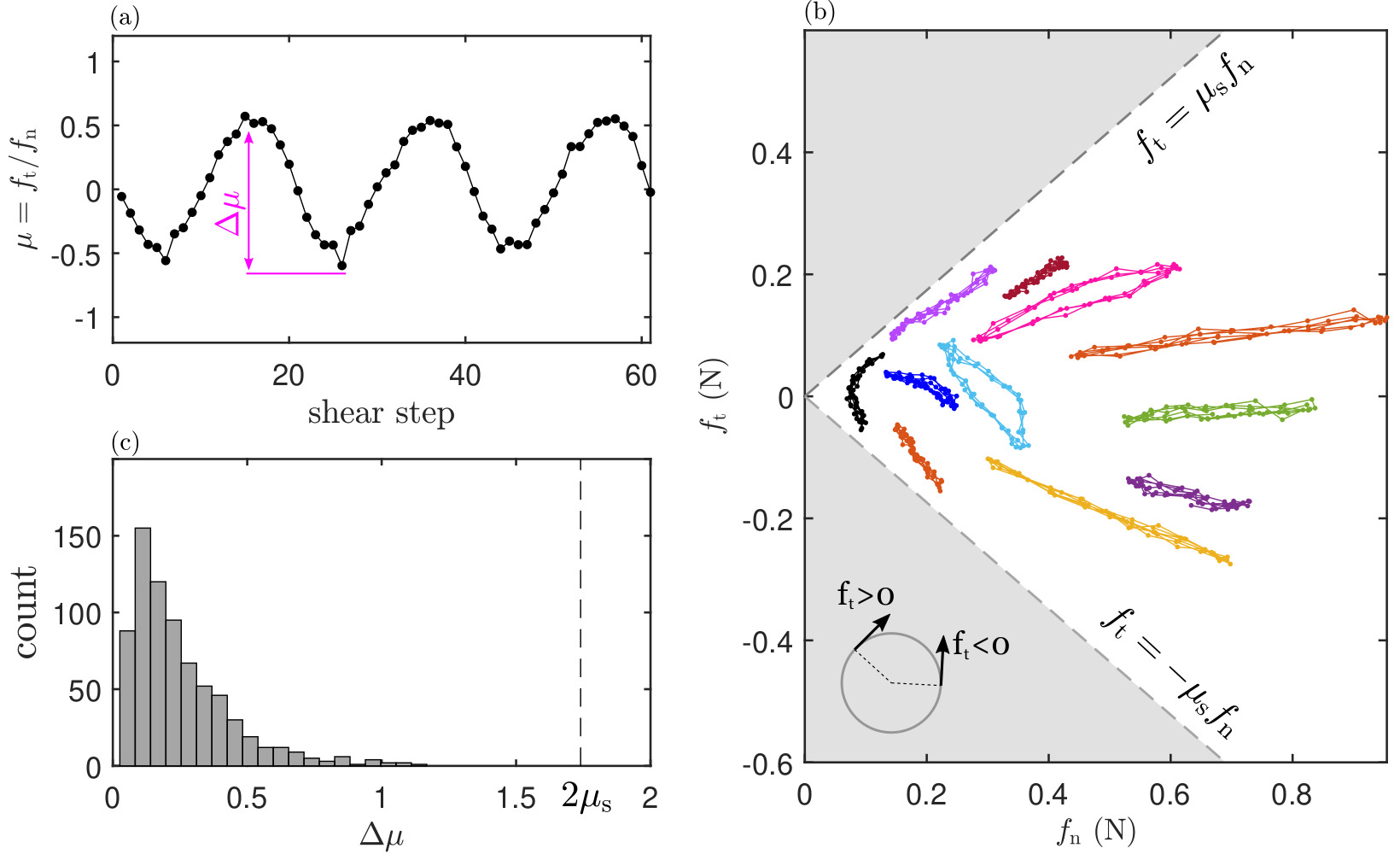}
    \caption{\yiqiu{ {\bf No evidence of sliding at persistent contacts.} All data in this figure are from the same ultrastable state as in Fig.~\ref{fig:non_persistent_contact}. (a) The evolution of $\mu=f_{\rm t}/f_{\rm n}$ for an example persistent contact over three consecutive shear cycles. (b) The evolution of forces on several example persistent contacts. Different colors represent different contacts. See supplementary video 2 for one example. The two dashed lines mark the conditions for the onset of sliding,
    where $\mu_{\rm s}=0.87$ is the static friction coefficient. The grey regions are inaccessible. The black data corresponds to the example contact shown in (a). The inserted schematic plots the sign convention for the tangential force components. (c) The number distribution of $\Delta\mu$ for all persistent contacts. The dashed line marks the value expected for a sliding contact.}} 
    \label{fig:persistent_contact_sliding}
\end{figure*}

\yiqiu{We find that there is no measurable sliding at the persistent contacts. 
The ultrastable states show the same strobed state over thousands of shear cycles, and the particles do not rotate from cycle to cycle, as can be seen from the supplementary videos of Ref.~\cite{zhao2022_prx}. The lack of rotation 
contrasts with recent 
experiments where particles do not move much but rotate significantly under cyclic loading, displaying contact sliding that leads to energy dissipation~\cite{Peshkov2019_PRE,Benson2022_PRL}. For an ultrastable state, there should be only two possible cases for a given persistent contact: (1) there is no sliding at contact; or (2) the two particles slide against each other during a shear cycle but return to the same position and stress states after a complete cycle. In the latter case, the tangential to normal force ratio $\mu = f_{\rm t}/f_{\rm n}$ should reach both $+\mu_{\rm s}$ and $-\mu_{\rm s}$ in a cycle, where $\mu_{\rm s}=0.87$ is the static friction coefficient of the particles. 
The evolution of $\mu$ on an example contact is shown in Fig.~\ref{fig:persistent_contact_sliding}(a). 
We measure the magnitude of variation of $\mu$, denoted as $\Delta\mu$ in Fig.~\ref{fig:persistent_contact_sliding}(a). 
\yiqiu{The example contact plotted in Fig.~\ref{fig:persistent_contact_sliding}(a) has the largest $\Delta\mu$ among all persistent contacts in the ultrastable state shown in Fig.~\ref{fig:persistent_contact_sliding}.}
In an ultrastable state, a contact that slides must have $\Delta\mu\approx 2\mu_{\rm s}$. (If sliding occurs in only one direction during the cycle, it would necessarily produce relative rotations of the two particles, which is not observed.) Figure~\ref{fig:persistent_contact_sliding}(c) plots the number statistics of $\Delta\mu$ for all persistent contacts in the example ultrastable state shown in Fig.~\ref{fig:non_persistent_contact}. Clearly, almost no persistent contact has $\Delta\mu$ near $2\mu_{\rm s}$, suggesting that these is no reversible sliding. We note that rolling without sliding is allowed and was observed at persistent contacts, but a complete characterization of rolling is beyond the scope of this paper. For completeness, we show the evolution of some example persistent contacts in the $(f_{\rm n},f_{\rm t})$ space in Fig.~\ref{fig:persistent_contact_sliding}(b).
\yiqiu{We did not perform the same analysis to non-persistent contacts because the measured $\mu$ becomes unreliable for weak forces due to the resolution limit of photoelastic force measurements. Extremely slow accumulation of plasticity induced by ratchet-like sliding on weak contacts, as found in numerical simulations~\cite{Alonso2004_PRL,McNamara2008_PRE}, cannot be completely ruled out.}}

The conditions that we used to identify persistent contacts and non-persistent contacts give high true positive ratio and a small true negative ratio. There are many contacts in our \yiqiu{system}
that do not satisfy either criterion. Whether these contacts are \yiqiu{persistent} or \yiqiu{non-persistent} could conceivably be resolved in future experiments with higher force, distance, and time resolutions. These unclassified contacts are not plotted in Fig.~\ref{fig:non_persistent_contact}(e).

\subsubsection{Contribution to the global elastic modulus}

\begin{figure}[!t]
    \centering
    \includegraphics[width =\columnwidth]{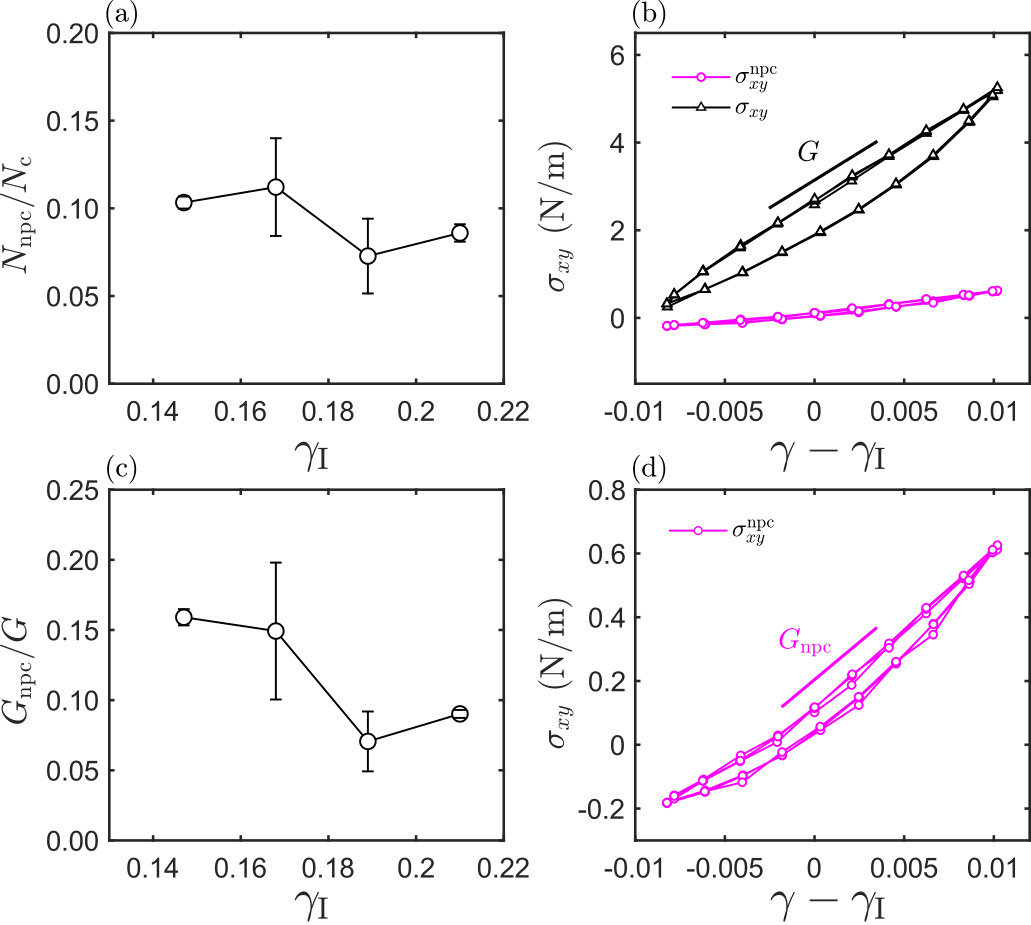}
    \caption{{\bf Statistics of the non-persistent contacts and their contributions to the emergent shear modulus.} 
    (a) The fraction of non-persistent contacts, defined as the number of non-persistent contacts divided by the total number of contacts, plotted for ultrastable states formed under same $\delta\gamma=0.95\%$ but different initial strains $\gamma_{\rm I}$. 
    Error bars are standard deviations computed from multiple ultrastable states with same $\gamma_{\rm I}$.
    (b) The shear stress contributed from all contacts, $\sigma_{xy}$, and the shear stress only from the non-persistent contacts, $\sigma_{xy}^{\rm npc}$ for an ultrastable state under three consecutive shear cycles.
    (c) The ratio between \yiqiu{$G_{\rm npc}$} and $G$ calculated from ultrastable states with
    \yiqiu{same} 
    $\delta\gamma=0.95\%$ but different $\gamma_{\rm I}$.
    Error bars are standard deviations computed from multiple ultrastable states with same $\gamma_{\rm I}$. 
    (d) shows a zoom-in to the purple curve ($\sigma_{xy}^{\rm npc}$) in (b).} 
    \label{fig:non_persistent_contact_statistics}
\end{figure}

We now show that the non-persistent contacts contribute a non-negligible amount to the emergent global elastic modulus. We calculate the shear stress contributed from the non-persistent contacts (npc) as 
\begin{equation}
    \sigma_{xy}^{\rm npc}= \yiqiu{\frac{1}{S}}\sum_{{\rm over~all~npc}~(i,j)}r_{ij,x}f_{ij,y}
\end{equation}
where the summation is only over all non-persistent contacts. Figure~\ref{fig:non_persistent_contact_statistics}(b) plots the total $\sigma_{xy}$ 
and $\sigma_{xy}^{\rm npc}$ for an example ultrastable state under cyclic shear. The contribution from non-persistent contacts to the shear modulus, $G_{\rm npc}$, is the slope of $\sigma_{xy}^{\rm npc}$, as sketched in Fig.~\ref{fig:non_persistent_contact_statistics}(d).  Figure~\ref{fig:non_persistent_contact_statistics}(c) plots the ratio $G_{\rm npc}/G$ for ultrastable states formed under $\delta\gamma=0.95\%$ but different $\gamma_{\rm I}$. We see that the contribution to the shear modulus from the non-persistent contacts can be as large as 10\% for $\gamma_I$ near the onset value for creating ultrastable states, and the actual contribution could be even larger because some unclassified contacts are likely non-persistent ones.
Thus, the non-persistent contacts make an appreciable contribution to the mechanical response of our ultrastable packings. \yiqiu{In addition, the importance of \yiqiu{the non-persistent contacts}
suggests that grains that might be identified as rattlers during some portion of the cycle may actually contribute to the elastic behavior observed for finite amplitude shear deformations. It also worth mentioning here that the scaling relation of Eq.~\ref{eq:G_p_scaling} is a global relation that contains contributions from all contacts. A detailed discussion of the separate contributions of \yiqiu{persistent}
and \yiqiu{non-persistent contacts}
to $G(p)$ is beyond the scope of the present paper.}

\subsection{Particle center trajectories}

Characterizing the particle motion within a shear cycle gives valuable insights into the nature of the mechanical responses of the packing. If an ultrastable packing deforms like a linear elastic continuum, all the particle displacements should define an affine deformation field. However, in our ultrastable states, the particles display clearly detectable, spatially correlated, non-affine displacements. In addition, some particle trajectories form loops with measurable enclosed area.
 
\begin{figure*}[!ht]
    \centering
    \includegraphics[width =\textwidth]{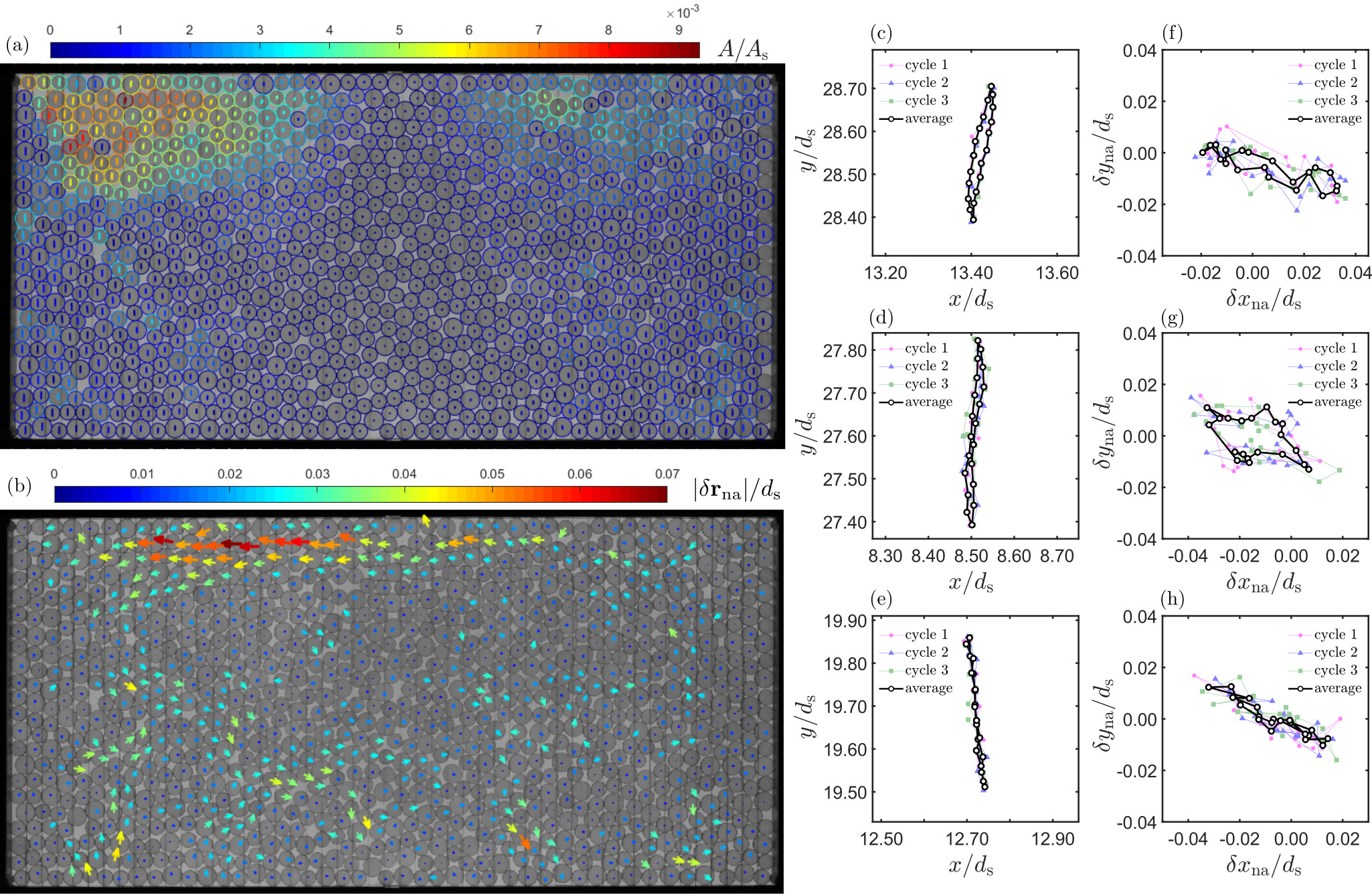}
    \caption{{\bf Particle center trajectories \yiqiu{display non-affine components and loops}.} \yiqiu{All data in this figure are from the same example} 
    ultrastable state as in Fig.~\ref{fig:non_persistent_contact}. (a) Particle center trajectories
    \yiqiu{averaged over three complete shear cycles plotted on top of an unpolarized image taken at the start of a cycle (state A in Fig.~\ref{fig:non_persistent_contact}(a)).}
    The color of trajectories indicates the normalized enclosed area of the trajectory $A/A_{\rm s}$, where $A_{\rm s}$ is the area of the small disc. Particle outlines are also colored as a guide to the eye. (b) The non-affine displacement field from state B to state D  in Fig.~\ref{fig:non_persistent_contact}(a) 
    \yiqiu{superimposed on  the same 
    unpolarized image  as in (a).}
    The arrow lengths are 20 times larger than the actual displacements, and colors indicate the displacement magnitudes
    \yiqiu{$|\delta\mathbf{r}_{\rm na}|/d_{\rm s}$}, where $d_{\rm s}$ is the diameter of the small disc. (c-\yiqiu{h}) Three example particle center trajectories     \yiqiu{(c-e}) and \yiqiu{their}
    corresponding non-affine trajectories (f-h) \yiqiu{calculated using Eq.~\ref{eq:non_affine}}. In (c-\yiqiu{h}), each data point \yiqiu{on the black curve} represents an average over three consecutive shear cycles in the ultrastable state. 
    \yiqiu{The trajectories for each of these cycles are also plotted.}}
    \label{fig:center_trajectories}
\end{figure*}

Figure~\ref{fig:center_trajectories}(a) shows all the particle center trajectories in a shear cycle for the example ultrastable state shown in Fig.~\ref{fig:non_persistent_contact}. The trajectories are nearly vertical lines that appear roughly consistent with an affine simple shear deformation field, suggesting that both the non-affine displacements and the enclosed loop areas are small. Figure~\ref{fig:center_trajectories}(b) plots the non-affine displacements of particles measured in the strain interval between state B to state D in Fig.~\ref{fig:non_persistent_contact}(a). Here, the non-affine displacement of particle $i$, $\delta \mathbf{r}_{{\rm na},i}$, is defined as
\begin{equation}\label{eq:non_affine}
 \delta\mathbf{r}_{{\rm na},i} = \delta\mathbf{r}_{i} - \frac{\sum_{j=1}^{N_p}\Theta(1.5
 \yiqiu{r}_{\rm s}- |x_i-x_j|)\delta\mathbf{r}_j}{\sum_{j=1}^{N_p}\Theta(1.5 
 \yiqiu{r}_{\rm s}- |x_i-x_j|)}    
\end{equation}
where $\Theta(x)$ is the Heaviside step function, $x_i$ is the x coordinate of particle $i$, $\delta\mathbf{r}_i$ is the the real displacement of particle $i$, 
$\yiqiu{r}_{\rm s}$
is the radius of the small particle, and $N_p$ is the total number of particles. In Fig.~\ref{fig:center_trajectories}(b), the arrows are colored according to the magnitude of the non-affine displacements $|\delta \mathbf{r}_{\rm na}|$, and the lengths of the arrows are 20 times $\yiqiu{|}\delta \mathbf{r}_{\rm na}\yiqiu{|}$. While the particles go back exactly to the same position after a full cycle, it is clear that their displacements within a shear cycle often contain significant non-affine components. \yiqiu{In the future, it could be interesting to compare our results to a recent theory which considers non-affine deformations while assuming no sliding at frictional contacts~\cite{Ishima2022_arxiv}.}

We further show that there are measurable loops formed by particle center trajectories, and also by the non-affine center trajectories. Figure~\ref{fig:center_trajectories} shows center trajectories 
\yiqiu{(c-e)} and 
\yiqiu{their corresponding}
non-affine center trajectories
\yiqiu{(f-h) calculated from Eq.~\ref{eq:non_affine}} for three example particles
\yiqiu{over three consecutive shear cycles}. 
\yiqiu{Note that the non-affine displacements appear noisier because they are near the accuracy of our particle center detection (about 0.01$d_{\rm s}$).}
The trajectory in (c) clearly \yiqiu{is a loop} and the one in (e) does not show measurable area. For the non-affine trajectories, only \yiqiu{(g)} shows a noticeable loop \yiqiu{above the noise level}. In numerical simulations of frictional \yiqiu{granular} systems, loops in particle trajectories~\cite{Royer2015_pnas} and in non-affine trajectories~\cite{Otsuki2021_EPJE} \yiqiu{were} observed. \yiqiu{In particular, the areas of these loops were found to obey a scaling relation with the elastic moduli of the system~\cite{Otsuki2021_EPJE}.} While relating these loops to global elastic responses is beyond the scope of the present paper, our work establishes their existence in this \yiqiu{experimental} frictional granular system.

\subsection{Stress correlations and emergent properties}

To obtain more insight on the stress responses of the ultrastable states, we consider the Vector Charge Theory of Granular mechanics and dynamics (VCTG)~\cite{nampoothiri2020_prl,nampoothiri2022_arxiv}. VCTG relates features of the stress correlations in the continuum limit to emergent elastic properties of the packing. We show that such a theory predicts forms of stress correlations that reasonably match our data. Notably, the results obtained from fitting the data to VCTG predictions uncover a feature that distinguishes the ultrastable states from the original shear-jammed states formed by initial shear alone.

\subsubsection{Defining ensembles with similar stress states}

\begin{figure*}[!t]
    \centering
    \includegraphics[width =\textwidth]{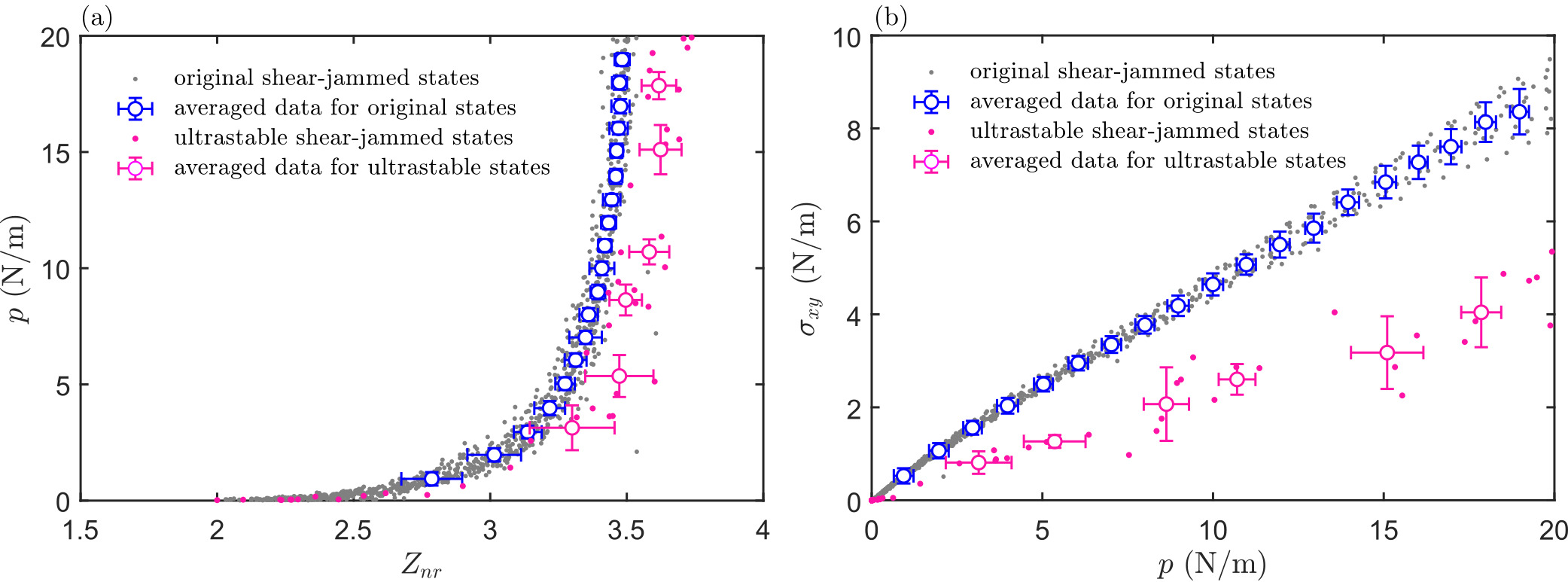}
    \caption{{\bf Group ultrastable states and original states into ensembles according to their stress states.}  (a) The pressure of the original shear-jammed states formed by initial shear alone (gray dots) and of the ultrastable states (purple dots) plotted versus the non-rattler contact \yiqiu{number}. The averaged values for states used in calculating stress correlations are also plotted. (b) The shear stress of the original shear-jammed states formed by initial shear alone (gray dots) and of the ultrastable states (purple dots) plotted versus pressure. The averaged values for states used in calculating stress correlations are also plotted.} 
    \label{fig:ensembles}
\end{figure*}

To obtain the correlation functions, we group ultrastable states with similar stress fields to form ensembles and calculate the averaged correlation functions over these ensembles. As shown in Fig.~\ref{fig:ensembles}(a) and (b), we show that all ultrastable states fall roughly on a same curve when plotting the non-rattler contact number $Z_{\rm nr}$ versus pressure $p$ or when plotting the shear stress $\sigma_{xy}$ versus $p$. This observation suggests that we group states according to $p$, and the states with similar $p$ will have similar $\sigma_{xy}$ and $Z_{\rm nr}$. Specifically, we group states with a pressure interval of $3$ N/m, and the averaged state variables for ultrastable states in these intervals are plotted using purple circles in Fig.~\ref{fig:ensembles}(a) and (b). The error bars mark the standard deviations.

For completeness, we also plot data from the original shear-jammed states that are formed by initial shear only in Fig.~\ref{fig:ensembles}(a) and (b). Comparing ultrastable states and original states provides additional insights into how cyclic shear modifies the mechanical properties of a jammed granular packing. Notably, for packings with similar $p$, ultrastable states usually contain more contacts and exhibit lower shear stress, suggesting that they are more stable and less anisotropic. We also group the original states according to intervals of pressure and apply the same stress correlation analysis below.

\subsubsection{Stress correlation functions}

We calculate the correlation functions between components of the stress tensor following the procedure detailed in Ref.~\cite{nampoothiri2020_prl} and Ref.~\cite{nampoothiri2022_arxiv}. 
We note that it is more convenient to examine the stress correlation functions in a reference frame $x'y'$ that is rotated 45 degrees from the original reference frame $xy$ (see Fig.~\ref{fig:protocol}). After the rotation, $x'$ is the principal dilation direction and $y'$ is the principal compression direction of the initial shear. The force chains in Fig.~\ref{fig:non_persistent_contact}(\yiqiu{g}) mostly align with direction $y'$. We consider below correlation functions and stress tensor expressions in this rotated frame.

The correlation functions in Fourier space are calculated as follows~\cite{nampoothiri2020_prl}
\begin{equation}\label{eq:correlation_def}
    C_{i'j'k'l'}(\mathbf{q}) = \langle \widetilde{\delta\sigma}_{i'j'}(\mathbf{q})\widetilde{\delta\sigma}_{k'l'}(-\mathbf{q})\rangle,
\end{equation}
where $\delta\sigma_{i'j'} = \sigma_{i'j'}-\langle\sigma_{i'j'}\rangle$ and $\langle\sigma_{i'j'}\rangle$ is the spatially averaged value in a packing and the $\langle\rangle$ in Eq.~\ref{eq:correlation_def} refers to average over different packings in an ensemble, and 
\begin{equation}
\widetilde{\delta\sigma}_{i'j'}(\mathbf{q}) =   \frac{1}{2\pi} \int     \delta\sigma_{i'j'}(\mathbf{r})e^{-i\mathbf{q}\cdot\mathbf{r}}d\mathbf{r}.
\end{equation}
We have used primed indices to emphasize that all the calculations are done in the rotated frame $x'y'$. 
More details on the calculation of the stress correlation functions are provided in the Supplementary Material~\footnote{See Supplementary Material for details on the correlation function calculations}. 

\begin{figure*}[!t]
    \centering
    \includegraphics[width = \textwidth]{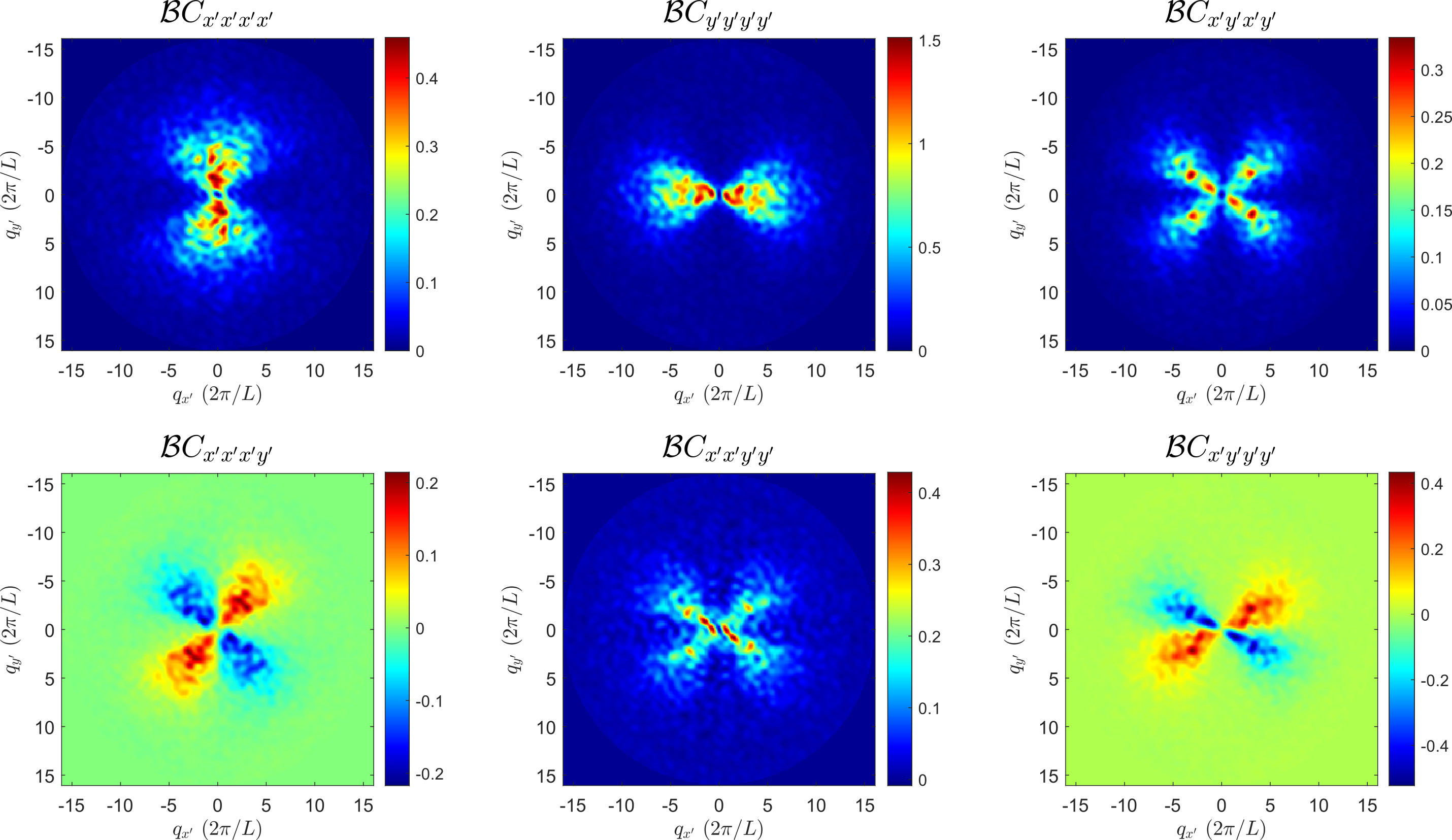}
    \caption{{\bf Stress correlation functions in the Fourier space averaged over 6 ultrastable states with similar stress states.}  The mean pressure for the 6 states used in the averaging process is $p=8.6\pm 0.7$ N/m. All the correlation functions are normalized by $\mathcal{B}$ from the VCTG fitting
    \yiqiu{(Eq.~\ref{eq:vctg_fit_form})}. Note that $C_{i'j'k'l'}(\mathbf{q})=\langle \widetilde{\delta\sigma}_{i'j'}\yiqiu{(\mathbf{q})}\widetilde{\delta\sigma}_{k'l'}\yiqiu{(-\mathbf{q})}\rangle$ by definition, and $x'$ and $y'$ are the principal compression and dilation directions of the initial simple shear strain field
    \yiqiu{(Fig.~\ref{fig:protocol}(a))}.} 
    \label{fig:stress-correlation-fourier-space}
\end{figure*}

As an example,  Fig.~\ref{fig:stress-correlation-fourier-space} shows the six stress-stress correlation functions in Fourier space obtained from an ensemble that contains 6 packings with a averaged pressure $p=8.6\pm 0.7$ N/m. We note that the general features of the correlation functions are consistent with those reported in Refs.~\cite{nampoothiri2020_prl,nampoothiri2022_arxiv}, including the pinch-point singularities at $|\mathbf{q}|\rightarrow 0$ and the obvious radial variation for wavelengths shorter than about $4d_{\rm s}$, where the continuuum theory is affected by the granularity of the medium. In Fig.~\ref{fig:stress-correlation-fourier-space} all correlation functions are normalized by $\mathcal{B}$, a 
parameter in the VCTG fitting form, which is presented below (Eq.~\ref{eq:vctg_fit_form}). 
Correlation functions for ultrastable states with other stress states share similar features.

\subsubsection{Elastic moduli from stress correlations}

\begin{figure*}[!t]
    \centering
    \includegraphics[width = \textwidth]{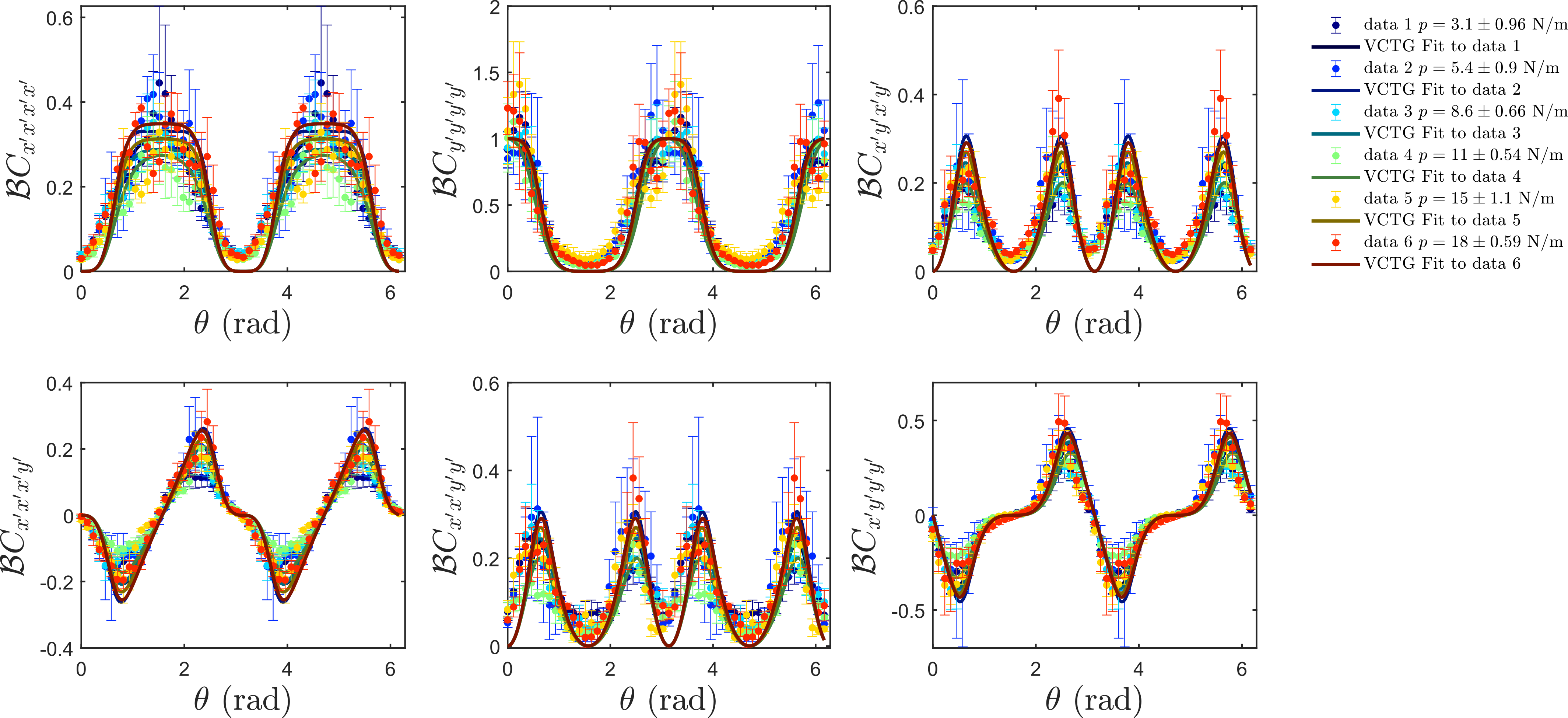}
    \caption{{\bf Angular variation of the stress correlations for ultrastable states in the long-wave-length limit and the fitted results from VCTG assuming an uniaxial symmetry.}  Correlation functions are normalized by the VCTG fitting parameter $\mathcal{B}$, whose variation with $p$ are shown in Fig.~\ref{fig:VCTG_fit_results}(a). The long-wavelength limit values are estimated by averaging data with $q$ between $2\pi/6d_{\rm s}$ and $2\pi/16d_{\rm s}$ where $d_{\rm s}$ is the diameter of the smaller disc particle in our packings.} 
    \label{fig:correlation_function_angular}
\end{figure*}

The emergent elastic moduli appear in the VCTG predictions of stress correlations, and we compute these by fitting the data. We first extract the angular dependence of the correlation functions in the long-wavelength limit and compare them to the VCTG predictions. Specifically, for each correlation function, we average the data in  a radial range between $2\pi/6d_{\rm s}$ and $2\pi/16d_{\rm s}$ and plot the radially averaged data versus the azimuthal angle $\theta$. Note that $16 d_{\rm s}$ is the size of the region of interest that we used to calculate these correlation functions, and $4d_{\rm s}$ is the length below which the correlation functions clearly deviates from the values at smaller $|\mathbf{q}|$. The radially averaged correlation functions are plotted in Fig.~\ref{fig:correlation_function_angular} for ultrastable states with different stress states as labeled by color. Note that each curve is averaged over several independent experimental realizations. We believe the scattering of data originates from that the number of packings used in the averaging process is rather small, and our system is not large enough. Nonetheless, we find that these curves can be reasonably characterized by the VCTG predictions.

\yiqiu{A key prediction from the VCTG theory is the form of the stress correlation functions in the long-wavelength limit~\cite{nampoothiri2020_prl}
\begin{equation}\label{eq:VCTG_general}
    C_{i'j'k'l'}(\mathbf{q}) = \epsilon_{i'a}\epsilon_{j'b}\epsilon_{k'c}\epsilon_{l'd}q_aq_bq_cq_d \langle \psi(\mathbf{q})\psi(\mathbf{-\mathbf{q}}) \rangle,
\end{equation}
where 
\begin{equation}\label{eq:vctg_psi}
    \langle \psi(\mathbf{q})\psi(\mathbf{-\mathbf{q}}) \rangle = \big(A_{i'j'}(\mathbf{q})\Lambda_{i'j'k'l'}A_{k'l'}(-\mathbf{q})\big)^{-1},
\end{equation} and 
\begin{equation}
    A_{i'j'} = q^2\delta_{i'j'}-q_{i'}q_{j'}.
\end{equation} Einstein notation applies to Eqs.~\ref{eq:VCTG_general}-\ref{eq:vctg_psi}, where $\epsilon_{i'j'}$ and $\delta_{i'j'}$ denote the Levi-Civita symbol and Kronecker delta. Detailed derivations of Eq.~\ref{eq:VCTG_general} can be found in Refs.~\cite{nampoothiri2020_prl,nampoothiri2022_arxiv}. Here, the only unknown variables are the elements of the 4-rank tensor $\Lambda$. These elements will be obtained by fitting the experimentally calculated stress correlation functions to Eq.~\ref{eq:VCTG_general}. 
}
In the VCTG framework, $\Lambda$ maps to the inverse elastic 
\yiqiu{constant}
tensor. \yiqiu{Using the symmetries of the elastic constant tensor $\Lambda_{ijkl} = \Lambda_{jikl}$ and $\Lambda_{ijkl} = \Lambda_{ijlk}$, there is
\begin{equation}
\begin{split}
   &\langle \psi(\mathbf{q})\psi(\mathbf{-\mathbf{q}}) \rangle =\\
   &(q_{y'}^4\mathcal{A}+q_{x'}^4\mathcal{B}+q_{x'}^2q_{y'}^2\mathcal{C}- q_{x'}^3q_{y'}\mathcal{D}-q_{x'}q_{y'}^3\mathcal{E})^{-1}
   \end{split}
\end{equation}
}
where 
\begin{equation}\label{eq:general_form_parameters}
    \begin{split}
        \mathcal{A} = & \Lambda_{1111} \\
        \mathcal{B} = & \Lambda_{2222} \\
        \mathcal{C} = & \Lambda_{1122} + 4\Lambda_{1212} + \Lambda_{2211}\\
        \mathcal{D} = & 2\Lambda_{1222} + 2\Lambda_{2212}\\
        \mathcal{E} = & 2\Lambda_{1112} + 2\Lambda_{1211}\\
    \end{split}
\end{equation}
For subscripts of $\Lambda$ we have used 1 and 2 to represent $x'$ and $y'$ for simplicity. Considerations of 
\yiqiu{additional} 
symmetries of the system \yiqiu{may lead to}
particular forms 
 \yiqiu{of the elastic constant tensor}
 that simplify the analysis~\cite{nampoothiri2020_prl,nampoothiri2022_arxiv}. In shear-jammed systems, it is reasonable to assume that the elastic moduli have uniaxial symmetry~\cite{Otto2003_pre}. Note that in these jammed states created by external stresses, the elastic moduli are determined by the geometry and topology of the force-bearing network that emerges from  the jamming process~\cite{nampoothiri2022_arxiv}.
In the Voigt form~\cite{Otto2003_pre}, such an elastic modulus tensor reads:
\begin{equation}
E
=\frac{1}{1-\nu_{x'}\nu_{y'}}\begin{pmatrix}
E_{x'} & \nu_{y'}E_{x'} & 0\\
\nu_{x'}E_{y'} & E_{y'} & 0 \\
0 & 0 & (1-\nu_{x'}\nu_{y'})G' \\
\end{pmatrix}.
\end{equation}
where $E_{x'}$
and $E_{y'}$
are the Young's moduli along $x'$, $y'$ directions, while
$\nu_{x'}$ and $\nu_{y'}$ stand for the Poisson's ratios along $x'$ and $y'$ directions. $G'$ is a shear modulus such that in the $x'y'$ coordinate system there is $\sigma_{x'y'} = G'\varepsilon_{x'y'}$. $G'$ is not the shear modulus $G$ which follows $\sigma_{xy} = G\gamma$. In addition, our imposed simple shear strain field has $\varepsilon_{x'y'} = 0$. Thus, $G'$ can not be extracted from our measured stress-strain curves.

The inverse elastic tensor then reads
\begin{equation}\label{eq:inverse_form1}
E^{-1}
=\begin{pmatrix}
1/E_{x'} & -\nu_{y'}/E_{y'} & 0\\
-\nu_{x'}/E_{x'} & 1/E_{y'} & 0 \\
0 & 0 & 1/G' \\
\end{pmatrix}.
\end{equation}
Note that this tensor maps to the $\Lambda$ tensor in the VCTG framework as following,
\begin{equation}\label{eq:inverse_form2}
E^{-1} \leftrightarrow \Lambda = 
\begin{pmatrix}
\Lambda_{1111} & \Lambda_{1122} & 2\Lambda_{1112}\\
\Lambda_{2211} & \Lambda_{2222} & 2\Lambda_{2212} \\
\Lambda_{1211} & \Lambda_{1222} & 2\Lambda_{1212} \\
\end{pmatrix}. \\
\end{equation}

Comparing Eq.~\ref{eq:inverse_form1} and Eq.~\ref{eq:inverse_form2} we get $\Lambda_{1211}=\Lambda_{1222}=\Lambda_{1112}=\Lambda_{2212}=0$. Thus, according to Eq.~\ref{eq:general_form_parameters}, we expect $\mathcal{D}=\mathcal{E}=0$ and 
\begin{equation}\label{eq:uniaxial_elastic}
\begin{split}
\mathcal{A} = & \Lambda_{1111} = \frac{1}{E_{x'}} \\
\mathcal{B} = & \Lambda_{2222} = \frac{1}{E_{y'}} \\
\mathcal{C} = & 4\Lambda_{1212} + \Lambda_{1122} + \Lambda_{2211} = \frac{2}{G'} -\frac{\nu_{y'}}{E_{y'}} - \frac{\nu_{x'}}{E_{x'}} \\
\end{split}
\end{equation}

Thus, we fit the correlation functions $C_{i'j'k'l'}(\theta)$ to the following form.
\begin{equation}\label{eq:vctg_fit_form}
\begin{split}
    C_{i'j'k'l'}(\theta) =  \frac{\epsilon_{i'a}\epsilon_{j'b}\epsilon_{k'c}\epsilon_{l'd}q_aq_bq_cq_d}{q_{y'}^4\mathcal{A}+q_{x'}^4\mathcal{B}+q_{x'}^2q_{y'}^2\mathcal{C}}
\end{split}
\end{equation}
Note that, we fit all six correlation functions together with three parameters $\mathcal{A}$, $\mathcal{B}$, and $\mathcal{C}$. In Fig.~\ref{fig:correlation_function_angular} we plot both the raw data and the fitted curves. All data and fitted curves are normalized by the fitting parameter $\mathcal{B}$ which depends on $p$. Notably, the fitted curves matches with the experimental data reasonably well. In addition, it also appears that most of the data collapse after the normalization by $\mathcal{B}$, suggesting roughly constant $\mathcal{A}/\mathcal{B}$ and  $\mathcal{C}/\mathcal{B}$ for ultrastable states with different $p$.

\subsubsection{Emergent elastic response}

The emergent elastic moduli, determined from fitting the measured stress-stress correlations to the VCTG predictions,  depend on  preparation protocols and the average stress state of a shaer-jammed solid. Here, we analyze the dependence of the three fitting parameters, $\mathcal{A}$, $\mathcal{B}$, and $\mathcal{C}$,  on the pressure $p$ of the  ultrastable states. We note that as the system unjams at $p=0$, what we are considering is scaling of the emergent properties near jamming. However, according to Eq.~\ref{eq:uniaxial_elastic}, while $\mathcal{A}$ and $\mathcal{B}$ can be directly related to the two Young's moduli, we can not extract the shear modulus and the Poisson's ratios from just the three fitting parameters. There are 5 unknowns but only three equations in Eq.~\ref{eq:uniaxial_elastic}. Ref.~\cite{nampoothiri2022_arxiv} demonstrates that additional equations can be obtained by considering material responses to additionally applied forces. In the present work, we report only scalings for the fitting parameters, and we leave the actual solutions for the shear modulus and Poisson's ratios to future investigations. 

\begin{figure*}[!t]
    \centering
    \includegraphics[width = \textwidth]{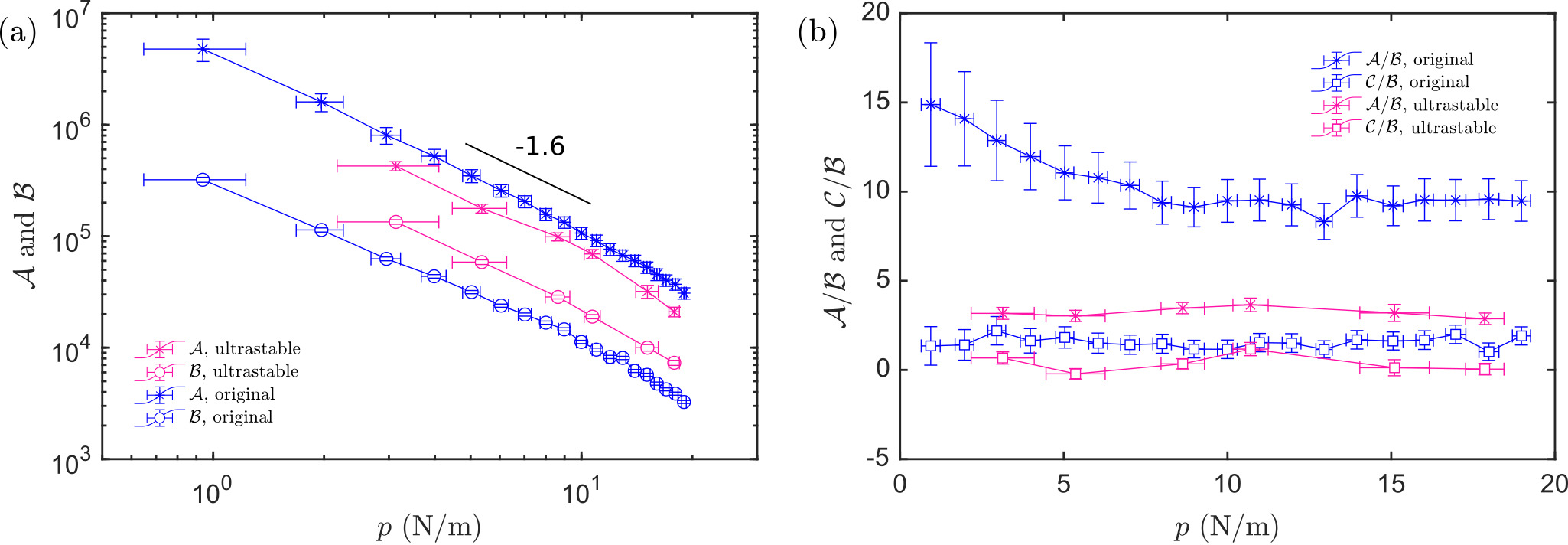}
    \caption{{\bf The VCTG fit results for the ultrastable states and the original states.} $\mathcal{A}$, $\mathcal{B}$, and $\mathcal{C}$ are defined in Eq.~\ref{eq:vctg_fit_form} (a) The VCTG fit results $\mathcal{A}$ and $\mathcal{B}$ plotted versus pressure for original and ultrastable states. Note that an interpretation is $E_{x'}=1/\mathcal{A}$ and $E_{y'}=1/\mathcal{B}$ where $E_{x'}$ and $E_{y'}$ are the Young's moduli of the anisotropic system along the principal dilation and compression directions of the initial shear respectively. (b) The ratios of the VCTG fit results $\mathcal{A}/\mathcal{B}$ and $\mathcal{C}/\mathcal{B}$ plotted versus pressure for original and ultrastable states. The ratio $\mathcal{A}/\mathcal{B}$ is clearly much larger for original states, constituting a feature to distinguish these two types of states. } 
    \label{fig:VCTG_fit_results}
\end{figure*}

We plot $\mathcal{A}$ and $\mathcal{B}$ as functions of $p$ in Fig.~\ref{fig:VCTG_fit_results}(a). While $\mathcal{A}>\mathcal{B}$ for all $p$, they both decay with increasing $p$. Notably, they appear to follow power laws with same exponent $\sim -1.6$. The divergence of these parameters suggest the Young's moduli vanish at the jamming point. We plot the ratios $\mathcal{A}/\mathcal{B}$ and $\mathcal{C}/\mathcal{B}$ in Fig.~\ref{fig:VCTG_fit_results}(b). Both ratios appear to be roughly constant. Notably, $\mathcal{A}/\mathcal{B}\approx 4$, meaning that the system is always stiffer along the $y'$ direction. It is interesting that $\mathcal{C}/\mathcal{B}\approx 0$. We do not yet have a clear understanding of this feature.

\subsubsection{A feature that distinguish ultrastable states from original shear-jammed states}

We show that the ratio $\mathcal{A}/\mathcal{B}$ can be used as a indicator to distinguish ultrastable states and the original shear-jammed states that are formed by initial shear alone. Figures~\ref{fig:VCTG_fit_results}(a) and (b) also show fit results obtained by performing same analysis on ensembles of the original shear-jammed states. Interestingly $\mathcal{A}$ and $\mathcal{B}$ follow power law scaling versus $p$ with same exponent as the ultrastable states. We find, however, that the ratio $\mathcal{A}/\mathcal{B}\geq 10$ for original shear-jammed states, which is much larger than for the ultrastable states. This means the original states have much more anisotropic elastic properties compared to the ultrastable states. 
We thus show that small-amplitude cyclic shearing changes the elastic response of a jammed packing. We emphasize that this observation is not equivalent to the changes of stress states as can be evidenced in Fig.~\ref{fig:ensembles}(b).  Instead,  it highlights that elasticity of these shear-jammed solids is truly an emergent phenomenon reflecting a rigidity that emerges from the complex interplay of local and global force and torque balance contstraints~\cite{nampoothiri2020_prl,nampoothiri2022_arxiv}.

\section{Concluding discussion}

In summary, we report a set of analyses on both global and local features of ultrastable
\yiqiu{shear-jammed granular materials}
in response to cyclic shear. We present three major findings.

First, we show that the emergent shear modulus $G$ for ultrastable states formed by different $\gamma_{\rm I}$ and $\delta\gamma$ falls on a single curve when plotted versus pressure $p$. A critical scaling near jamming between $G$ and $p$ is examined extensively in numerical simulations~\cite{OHern2003_pre,somfai2007_pre,vanderwerf2020_prl,Ishima2020_PRE,Morse2021_PNAS,pan2022_arxiv}, and is of key interests in the scaling theories of jamming~\cite{Goodrich2016_pnas}. 
Notably, \yiqiu{the ultrastable states follow} $G\sim p^{\beta}$ with $\beta\approx 0.5$,
\yiqiu{consistent with a numerical simulation with particles having similar friction coefficient and contact force law~\cite{somfai2007_pre}}. 
To our knowledge, the range of boundary strain within which 
\yiqiu{a frictional} 
system behaves elastically \yiqiu{is usually very small
because boundary strain may induce sliding at contacts~\cite{dominic2000_jgge,alonso2005_pre,Otsuki2017_PRE,sun2022_soft}.} 
\yiqiu{Thus the shear modulus has typically been determined from measurements of sound speeds~\cite{makse2004_pre}
rather than from stress-strain curves.
Consistent with this picture, the original shear-jammed states in our experiments were observed to deform plastically under any given cyclic strain amplitude~\cite{zhao2022_prx}. \yiqiu{As previously reported,} highly elastic ultrastable states emerged under cyclic shearing, which induces changes in the distribution of friction forces at contacts~\cite{zhao2022_prx}.} 
\yiqiu{We have now found that there is no measurable 
sliding for the persistent contacts that carry the majority of the forces in these ultrastable states.}

Second, we \yiqiu{find nontrivial grain scale motions within a shear cycle in an ultrastable state. A measurable fraction of contacts} open and close reversibly \yiqiu{during a cycle, and these contacts make a} non-negligible \yiqiu{contribution} to the emergent elastic modulus. \yiqiu{Thus} predictions based on a contact network with a fixed geometry presumably cannot completely account for the macroscopic elasticity of these states. \yiqiu{It is known that} 
the distribution of small inter-particle gaps and weak contact forces are intimately connected to packing stability~\cite{Charbonneau2015_prl,Babu2022_PRE,wang2022_pnas}. Our work demonstrates \yiqiu{that} reversible activation of these gaps may lead to non-trivial dynamical phases in a frictional, shear-jammed system.
\yiqiu{We also observe non-affine particle displacements, with some particles moving around loops with finite enclosed area.
It would be interesting to compare the observed particle displacement fields to the low-frequency vibrational modes that can be calculated from the experimental data~\cite{Zhang2017_natcom}, where one may find analogies to features found in model glasses, such as string-like dynamical defects~\cite{Hu2022_Natphys}.}

Third, we examine the relation between the spatial stress fluctuations and the emergent elastic constants of the ultrastable states from the perspective of the Vector Charge Theory of Granular mechanics and dynamics (VCTG)~\cite{nampoothiri2020_prl,nampoothiri2022_arxiv}. 
In the long-wavelength limit, 
\yiqiu{the stress-stress correlation functions measured from ultrastable and original shear-jammed states}
matches well with the predictions by VCTG for an anisotropic \yiqiu{system} 
with uniaxial symmetry. 
Fitting our data to the theory, we extract the values of three parameters.  Two of these are the Young's moduli $E_{x'}$ and $E_{y'}$, and the third is a linear combination of the  Poisson ratios $\nu_{x'}$, $\nu_{y'}$ and a shear modulus $G'$.
Note that $x'$ and $y'$ are the principal compression and dilation directions of the initial shear. 
We find that, for both original shear-jammed states and the ultrastable states, $E_{x'}$ and $E_{y'}$ scale as a power-law with pressure $p$, sharing same exponent $\alpha\approx 1.6$. 
The vanishing of the Young's moduli
as $p\rightarrow 0 $ is 
\yiqiu{consistent qualitatively with the vanishing of the shear modulus $G$ measured independently from the stress-strain curves.} The relation between \yiqiu{the} exponent 
\yiqiu{$\alpha$} and the exponent $\beta$ that links $G$ and $p$ is an interesting topic for further investigation. \yiqiu{We note that the elastic moduli obtained from VCTG fittings are linear elastic constants, while  stress-strain curves contain contributions from non-linear features like the non-persistent contacts.}
The ratio $E_{x'}/E_{y'}$ is always at least twice as large for original shear-jammed states as for the ultrastable states,
suggesting that small-amplitude cyclic shearing significantly alters the elastic properties of a jammed packing. 
In addition, the ratio 
\yiqiu{$E_{x'}/E_{y'}$} 
does not approach 1 as $p\rightarrow 0$, suggesting that the system remains anisotropic at jamming
\yiqiu{point, which is a feature of the shear jamming transition~\cite{Bi2011_nat,Baity2017_jps,chen2018_pre,Xiong2019_gm}}.
Additional experiments and analysis probing the system response to a point force~\cite{nampoothiri2022_arxiv} may help to determine the Poisson ratios and the shear modulus $G'$.

\vspace{20pt}

\noindent \textbf{Author contributions}

\vspace{5pt}

YiZ, BC, and JS contributed to conception and design of the study and collaborated on the interpretation of the experimental results. YiZ, YuZ, DW, and HZ designed the experimental procedure and calibrated the apparatus. YiZ and YuZ conducted the experiments. YiZ performed the data analysis. YiZ, BC, and JS wrote the first draft of the manuscript. All authors contributed to manuscript revision, read, and approved the submitted version.

\vspace{10pt}

\begin{acknowledgments}
YiZ thanks Yinqiao Wang for helpful discussions on calculating the correlation functions.
YiZ thanks Peter K. Morse, Shuai Zhang, Yuliang Jin, and Deng Pan for helpful discussions about the scaling of shear modulus.
This work was primarily supported by NSF grant DMR-1809762. BC was supported by NSF grants CBET-1916877, and CMMT-2026834, and BSF-
2016188. 
\end{acknowledgments}

\appendix

\section{Calculation of $C_{ijkl}(\theta)$ in the long-wavelength limit}

\subsection{The rotated reference frame $x'y'$}

As mentioned in the main text, we find that it is most convenient to examine the stress correlations in a frame $x'y'$ that is rotated by $\pi/4$ clock-wisely from the original $xy$ frame. We express everything in this rotated frame. We show in Fig.~\ref{fig:SI_1}(a) a polarized image in the $x'y'$ frame. The axes for the original $xy$ directions are also plotted. 

\subsection{Construct stress fields}

Following the procedure detailed in Ref.~\cite{nampoothiri2020_prl}, we first define a particle-scale stress tensor for each individual disc. For the $g$th disc, we define
\begin{equation}
    \hat{\sigma}_{g} = \frac{1}{A_g}\sum_{k=1}^{z_g}\mathbf{r}_{k}\otimes\mathbf{f}_k,
\end{equation}
where $A_g$ is the Voronoi area for the $g$th particle, $\mathbf{r}_k$ and $\mathbf{f}_k$ are the branch vector and contact force vector corrspond to the $k$th contact, and the summation of $k$ goes over all contacts on the $g$th particle. 

To construct a 
stress field $\sigma_{ij}$, where $i$ and $j$ can be either $x'$ or $y'$, we let $\sigma_{ij}(x',y') = \hat{\sigma}_{g,ij}$ if $(x,y)$ belongs to the Voronoi cell of particle $g$. Figure~\ref{fig:SI_1}(b) shows the constructed $\sigma_{y'y'}$ field for the state shown in Fig.~\ref{fig:SI_1}(a).

\subsection{Calculate the Fourier transformation of stress fields}

\begin{figure*}[!ht]
    \centering
    \includegraphics[width = \textwidth]{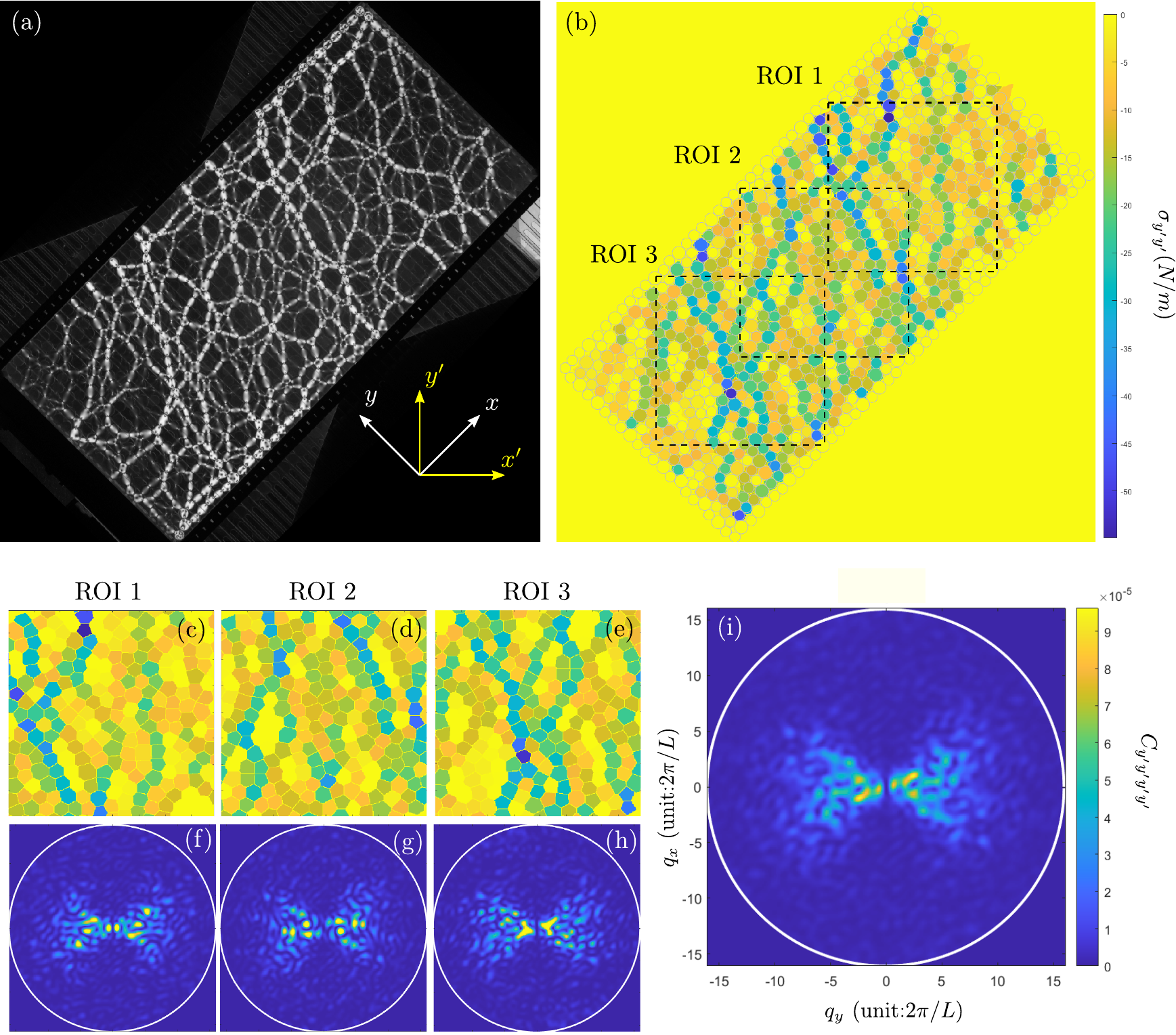}
    \caption{(a) The plolarized image for an example ultrastable state formed by $\gamma_{\rm I} = 0.147$ and $\delta\gamma = 0.95\%$. The direction of axes are sketched. (b) The continuous stress field $\sigma_{y'y'}$ constructed from the particle-scale stress tensors for the state shown in (a). The three regions of interest (ROI) are sketched. (c-e) shows zoom-in versions of the three ROIs in (b). (f-g) shows the correlation functions $C_{y'y'y'y'}$ calculated from the stress fields in (c-e) respectively. (i) The correlation function for the state in (a) obtained by averaging the correlation functions shown in (f-g). The color scales in (c-e) are the same as in (b). The color scales in (f-h) are the same as in (i). Note that $L=16d_{\rm s}$ where $d_{\rm s}$ is the diameter of the smaller disc. $L$ is the side length of the square ROIs.}
    \label{fig:SI_1}
\end{figure*}

To avoid complications introduced from the system boundaries, we consider three square regions of interest (ROIs) as shown in Fig.~\ref{fig:SI_1}(b). Each ROI has a side length $L=16d_{\rm s}$ where $d_{\rm s}$ is the diameter of the smaller disc.

\begin{figure*}[!t]
    \centering
    \includegraphics[width = \textwidth]{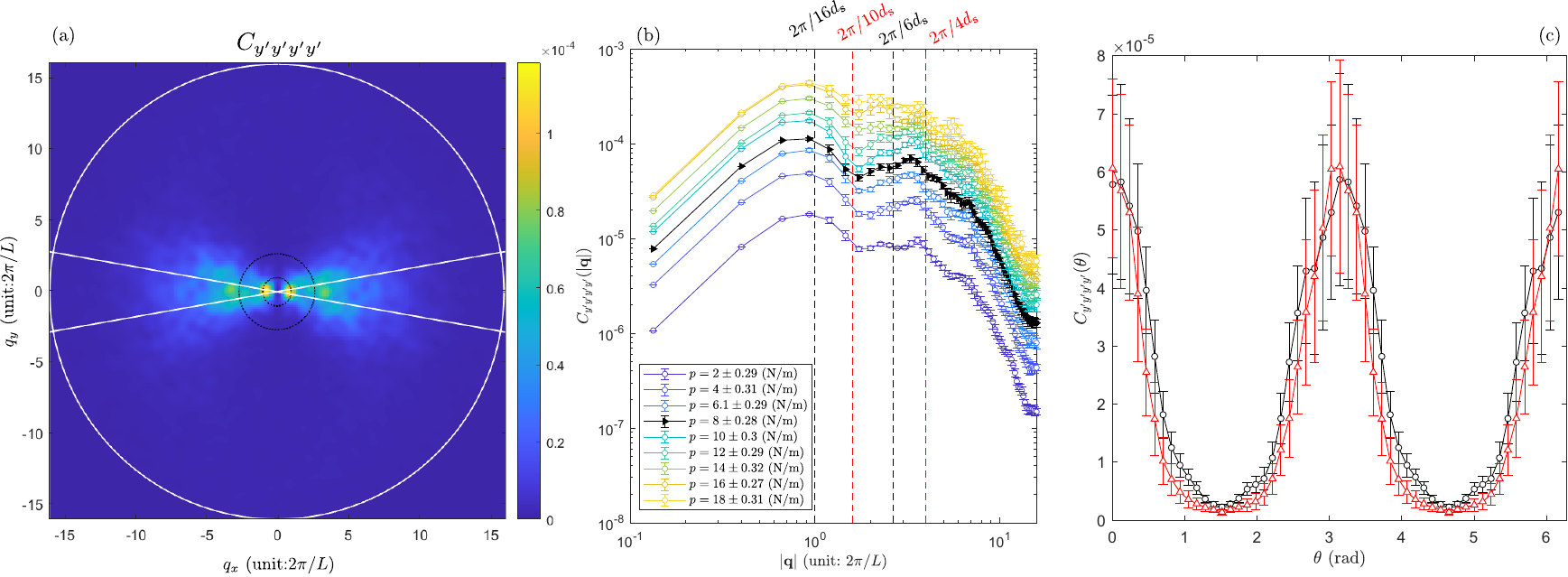}
    \caption{(a) The ensemble-averaged correlation function $C_{y'y'y'y'}(\mathbf{q})$ from 30 original shear-jammed states with averaged pressure $p=8\pm 0.28$ N/m. The two black dotted circles have radii $2\pi/16d_{\rm s}$ and $2\pi/6d_{\rm s}$. The two white lines form an angle $\pi/9$, within which the radial functions $C_{y'y'y'y'}(|\mathbf{q}|)$ shown in (b) are calculated. (b) The radial dependence of the correlation functions for ensembles of original shear-jammed states with different stress states. The black curve corresponds to the function shown in (a). (c) The angular function $C_{y'y'y'y'}(\theta)$ for the same ensemble shown in (a) calculated by averaging radial parts within the regime between the two black dashed lines in (b) (black circles) and within the regime between the two red dashed lines in (b) (red triangles).}
    \label{fig:SI_2}
\end{figure*}

For each ROI stress field $\sigma_{ij}$, we first calculate the deviation from the mean 
\begin{equation}
    \delta\sigma_{ij}(x',y') = \sigma_{ij}(x',y') - \langle \sigma_{ij} \rangle,
\end{equation} 
where $\langle\cdot\rangle$ represent a spatial average. We then calculate the Fourier transform of $\delta\sigma_{i'j'}$ as following

\begin{equation}
    \widetilde{\delta\sigma}_{ij}(\mathbf{q}) = \frac{1}{2\pi}\int\delta\sigma_{ij}(\mathbf{r})e^{-i\mathbf{q}\cdot\mathbf{r}}d\mathbf{r}
\end{equation}

In practice, we perform the discrete Fourier transformation
\begin{widetext}
\begin{equation}
    \widetilde{\delta\sigma}_{ij}(q_{x'},q_{y'}) = \frac{1}{2\pi}\sum_{n = 1}^{N}\sum_{m=1}^{N}\delta\sigma_{ij}(n\Delta {x'},m\Delta {y'})e^{-iq_{x'}n\Delta {x'}}e^{-iq_{y'}m\Delta {y'}}\Delta {x'}\Delta {y'},
\end{equation}
\end{widetext}
where $\Delta x' = \Delta y' =td_{\rm s}/2$ and $t=0.81$ is a constant. We have tested that using a smaller $t$ gives similar results. $N\Delta x'=N\Delta y' = L$ that is the side length of the ROIs.

\subsection{Calculate the Fourier transform of the correlation functions}

The stress correlation functions in the Fourier space for a certain packing is calculated by
\begin{equation}
    C_{ijkl}(\mathbf{q}) = C_{ijkl}(q_{x'},q_{y'}) = \widetilde{\delta\sigma}_{ij}(q_{x'},q_{y'})\widetilde{\delta\sigma}_{kl}(-q_{x'},-q_{y'}).
\end{equation}

For each packing, we first calculate $C_{ijkl}$ for the three ROIs, and the stress-stress correlation function $C_{ijkl}$ for this state is obtained by averaging over the three ROIs. Figure~\ref{fig:SI_1}(f-h) plots the calculated $C_{y'y'y'y'}$ for the three ROIs as shown in (c-e), while the final result is shown in (i) which is obtained by averaging the three functions shown in (f-h).

\subsection{Calculate the ensemble averaged correlation functions}

After obtaining the correlation function $C_{ijkl}(\mathbf{q})$ for all states. We calculate the ensemble-averaged correlation functions for states with similar stress states. The final result is              
\begin{equation}
    C_{ijkl}(\mathbf{q}) = \langle C_{ijkl}(\mathbf{q}) \rangle,
\end{equation}
where $\langle\cdot\rangle$ represents ensemble average. In the main text, we always consider the ensemble-averaged correlation functions. Figure~\ref{fig:SI_2}(a) plots the ensemble averaged correlation function $C_{y'y'y'y'}$ over 30 different packings of original shear-jammed states, which appears more smooth than correlation functions obtained from individual packings such as the one shown in Fig.~\ref{fig:SI_1}(i).

\subsection{Identify the continuum limit using the radial dependence of the correlation functions}

The Vector Charge Theory of Granular Mechanics (VCTG)~\cite{nampoothiri2020_prl,nampoothiri2022_arxiv} predicts features of the correlation functions in the long-wavelength limit ($|\mathbf{q}|\rightarrow 0$). To compare to the theory, we need to identify the range of $|\mathbf{q}|$ where the system can be regarded to be in the long-wavelength limit. Thus, we examine the radial dependence of the correlation functions in our systems and focus on the regime where $C_{ijkl}$ does not depend on $|\mathbf{q}|$.

As examples, we consider ensembles of original shear-jammed states. We plot $C_{y'y'y'y'}(|\mathbf{q}|)$ along $\theta=\pi$ direction in Fig.~\ref{fig:SI_2}(b) for ensembles with different stress states. In practice, these curves are averaged from angular direction $\theta\in (\pi - \pi/18,\pi + \pi/18)$. The black triangle curve corresponds to the correlation function shown in (a). The data with $|\mathbf{q}|<2\pi/L$ is not of our interest. All $C(\theta)$ data shown in the main text are obtained by averaging over $|\mathbf{q}|$ between $2\pi/16d_{\rm s}$ and $2\pi/6d_{\rm s}$, as marked by the two black dashed lines in Fig.~\ref{fig:SI_2}(b). In this regime, $C_{y'y'y'y'}$ roughly display a plateau expect perhaps near the system size ($|\mathbf{q}|\approx 2\pi/16d_{\rm s}$) where it shows a clear growth. A similar feature was carefully examined in Ref.~\cite{lemaitre2021_prl} and was attribute to the history-dependent nature of frictional contact forces. The plateau is perhaps better defined in the regime between $2\pi/10d_{\rm s}$ and $2\pi/4d_{\rm s}$ as marked by the two red dashed lines in Fig.~\ref{fig:SI_2}(b). We find that averaging in the regime between the two black dashed lines or between the two red dashed lines do not lead to drastically different angular functions $C_{y'y'y'y'}(\theta)$. For example, the black and red curve in Fig.~\ref{fig:SI_2}(c) are correlation functions for same ensemble as shown in (a) whose radial parts are averaged in the regime between the two black dashed lines and in the regime between the two red dashed lines respectively. It appears that the angular dependence of the two curves remain almost the same.

\subsection{Calculate the angular dependence of the cross-correlation functions}

After identifying the range of $|\mathbf{q}|$ where $C(\mathbf{q})$ displays a plateau, we calculate the angular functions $C(\theta)$ by averaging the radial parts in the plateau regime. In the main text, all $C({\theta})$ curves are calculated by averaging the radial parts in the regime between the two black dashed lines in Fig.~\ref{fig:SI_2}(b).


\bibliography{main}

\end{document}